# Electron Localization Enhances Cation Diffusion in Transition Metal Oxides: An Electronic Trebuchet Effect

Yanhao Dong[1], Liang Qi[2], Ju Li[1,3], and I-Wei Chen[4*]

[1]Department of Nuclear Science and Engineering, Massachusetts Institute of Technology, Cambridge, MA 02139, USA

[2]Department of Materials Science and Engineering, University of Michigan, Ann Arbor, MI 48109, USA

[3]Department of Materials Science and Engineering, Massachusetts Institute of Technology, Cambridge, MA 02139, USA

[4]Department of Materials Science and Engineering, University of Pennsylvania, Philadelphia, PA 19104, USA

[*]**Corresponding Author Information**

**Tel: +1-215-898-5163; Fax: +1-215-573-2128**

*E-mail address*: <u>iweichen@seas.upenn.edu</u> (I-Wei Chen)

*Postal address*: Department of Materials Science and Engineering, University of Pennsylvania, LRSM Building, Room 424, 3231 Walnut St., Philadelphia, PA 19104-6272

**Abstract**

Ion diffusion is a central part of materials physics of fabrication, deformation, phase transformation, structure stability and electrochemical devices. Conventional theory focuses on the defects that mediate diffusion and explains how their populations influenced by oxidation, reduction, irradiation and doping can enhance diffusion. However, we have found the same influences can also elevate their mobility by orders of magnitude in several prototypical transition-metal oxides. First-principles calculation fundamentally connects the latter observation to migrating ion's local structure, which is inherently soft and has a broken symmetry, making it susceptible to electron or hole localization, thereby realizing a lower saddle-point energy. This finding resolves an unanswered question in physical ceramics of the past 30 years: why cation diffusion against the prediction of classical nonstoichiometric defect physics is enhanced in reduced zirconia, ceria and structurally related ceramics? It also suggests the saddle-point electron-phonon interaction that enables a negative-$U$ state is akin to the counterweight effect that enables a trebuchet. This simple picture for the transitional state explains why enhanced kinetics mediated by radical-like-ion migration occurs often, especially under extreme conditions.

Atomic diffusion is a fundamental process that enables changes of shape, composition and structure, and is thus important for many applications. This includes standard materials problems such as fabrication, microstructure, functionality and stability. (*1-4*) But it also includes emergent materials problems especially those involving extreme conditions, e.g., field-assisted sintering/processing (*5*, *6*), turbine blade oxidation (*7*), high-energy-density batteries(*8*, *9*), neuromorphic memristor switching (*10*), photovoltaic inorganic-organic halide perovskites cells (*11*) and irradiation effects in transmission electron microscopy and nuclear reactors. In a crystal, the elementary step in diffusion is the movement of an atom from one lattice site to another through a non-lattice-site saddle point, which has a very different local environment and symmetry. Naturally, the lattice site and the saddle point must render very different electronic configurations to the atom, and such changes in electronic configurations can profoundly affect the transition probability. However, the conventional theory has placed a greater focus on the availability of a neighboring vacant site, itself a lattice defect with which the lattice atom exchanges positions during diffusion. While the theory elaborates on how the defect population may be tuned by doping and other external means, it mostly ignores the possibility that the same tuning may affect the electronic configuration of the saddle-point state, thus the defect mobility. This mobility aspect is the subject of the present work. Using first-principles calculations, we will evaluate how the electronic configuration and energetics of the saddle-point state are altered by oxidation and reduction, which are commonly encountered during material processing and applications. The result helps us understand why defect mobility hence atomic diffusivity is so sensitive to the environment, especially ones provided by extreme conditions.

In practical terms, the question we will address is the following: is diffusion faster or slower in a reduced oxide than in an unreduced one, and why? This is an interesting question because reduction-generated oxygen vacancies and electrons have opposite effects on ion diffusion, for example that of cation. First, concerning the mobility, while a reduced cation with a lower valence should experience less electrostatic interaction and diffuse faster, it also has a larger ionic size which hampers diffusion. Second, concerning the defect population, more oxygen vacancies will leave fewer cation vacancies according to the thermodynamics of the Schottky pair, which comprises a stoichiometric set of cation vacancy and anion vacancy. Thus, cations that move by exchange with neighboring cation vacancies will diffuse less. These competing effects are further complicated by aliovalent dopants—and their charge-compensating defects—that are often added to transition metal oxides to increase their technological appeal.

Recently, we have demonstrated that atmospheric and electrical reduction can enhance cation diffusion by many orders of magnitude. *(12-14)* This observation was first made on zirconia and ceria *(12-21)*, which have a fluorite structure and are widely used for fast oxygen conductors. But it is also corroborated in $ABO_3$ ($B$ = transition metal) perovskites *(21-25)* that host several fast oxygen conductors. Now that the experimental fact is clear, this study will answer why, by undertaking a first-principles study to unravel the mechanism of reduction-enhanced cation diffusion. Much is already known about the electronic and defect structure of these oxides, and vacancy exchange has indeed been verified as the diffusion mechanism for cations. *(26-29)* This knowledge guided our calculations to focus on cubic $ZrO_2$, $CeO_2$ and $BaTiO_3$ with several specific lattice/electronic defect configurations. Below we provide a brief synopsis of the results.

We found oxygen vacancy and electron localization synergistically lower cation's migration barrier over the saddle point. Energetically, the lowered barrier is mostly attributed to a strong tendency for electron localization on the migrating cation at the saddle point. This is reminiscent of a negative-$U$ state (*30*) and is made possible by the soft saddle-point environment that enables a static, strongly stabilizing electron-phonon interaction. The localizing electron may be viewed as providing a kind of electronic trebuchet effect, in that the extra electron experiencing a decrease in energy helps a cation migrate much like a counterweight experiencing a decrease in potential energy helps a projectile launch. Our study has also provided a set of striking examples of how mass transport of the slowest-moving species is profoundly influenced by highly mobile though unrelated species, e.g., oxygen vacancy and electron that are unrelated to cations can nevertheless influence cation diffusion, which is rate-limiting in all these oxides. These results are not only relevant to cation diffusion in other transition metal oxides of a similar band gap (~3.5 eV in $BaTiO_3$ to ~5 eV in $ZrO_2$) and electronic configuration, but also to enhanced anion diffusion for which we have found a similar mechanism mediated by "radical-like" species with hole localization. Therefore, the electronic trebuchet mechanisms for the transitional states are quite general and can explain many observations of enhanced kinetics under common and extreme conditions in a broad range of applications.

**Reduction lowers cation migration barrier in $ZrO_2$ and $CeO_2$**

We studied cation ($M$=Zr, Ce) exchange with its vacancy at (½½0) in four models (**Fig. 1a-b**): Model A, $M^{4+}$ without a (neighboring) $V_O^{\cdot\cdot}$; Model B, $M^{3+}$ (including the extra electron generated by reduction) without a $V_O^{\cdot\cdot}$; Model C, $M^{4+}$ with a $V_O^{\cdot\cdot}$; and Model D, $M^{3+}$ with a $V_O^{\cdot\cdot}$. Here, we allow the possibility of having a neighboring $V_O^{\cdot\cdot}$ at the exchange site because a slow diffusing $M$ cation has plenty time to wait for a fast diffusing $V_O^{\cdot\cdot}$ to arrive before attempting a successful exchange. The calculated trajectories and energies defining the saddle point configurations and energy barriers are summarized in **Fig. 2c**, **Fig. S1-S2** (for saddle point configurations) and **Table S1**.

The size effect of $V_O^{\cdot\cdot}$ is seen by comparing Model A/B with Model C/D. First, with two interfering O in Model A and B, the trajectory veers into a cation-absent neighboring cell along the red dashed curve in **Fig. 1a** to minimize short-range repulsion. In contrast, with one interfering oxygen ion removed in Model C and D, a straight-line trajectory along the black dashed curve in **Fig. 1a** is followed. Second, at the saddle point, $M^{4/3+}$ is surrounded by six O in Model A and B (**Fig. 1a**, upper inset; the two interfering O marked by black arrows are closer than the other four), and by five O in Model C and D (**Fig. 1a**, lower inset; the one marked by the arrow is closer). Third, $V_O^{\cdot\cdot}$ lowers the migration barrier by 0.33/0.73 eV in $ZrO_2$ from Model A/B to C/D, and by 0.47/1.51 eV in $CeO_2$ from Model A/B to C/D. Therefore, $V_O^{\cdot\cdot}$ helps cation migration.

The charge effect of cation reduction is seen by comparing Model A/C with Model B/D. First, the extra electron is localized at the target (migrating) cation in both the ground state and the saddle-point state, as indicated by the newly formed $M$ $d/f$ state below the Fermi level (see the calculated projected density of state—DOS—of the target cation in **Fig. 1d-e** and **Fig. S4**, **S6, S8** and **S10**). This causes an increased Bader charge *(32)* and a longer *M*-O distance because a reduced cation $M^{3+}$ has a larger radius than $M^{4+}$, as summarized in **Table S1**. Second, the migration barrier is lowered in three out of four cases: by 0.74/1.14 eV in $ZrO_2$ from Model A/C to B/D, and by 0.68 eV in $CeO_2$ from Model C to D. Therefore, cation reduction also helps cation migration.

The two effects are synergistic in that their combination lowers the barrier more than their sum. For example, the lowering from Model A→D (1.47 eV for $ZrO_2$ and 0.94 eV for $CeO_2$) is much larger than the sum from A→B (the charge effect due to one extra electron alone) and A→C (the size effect due to one $V_O^{\cdot\cdot}$ alone), which is 1.07 eV for $ZrO_2$ and −0.11 eV for $CeO_2$. Together, they lower the migration barrier to 3.17 eV for $Zr^{3+}$ in Model D from 4.64 eV for $Zr^{4+}$ in Model A, and to 3.28 eV for $Ce^{3+}$ from 4.32 eV for $Ce^{4+}$.

To understand why synergism can greatly enhance cation mobility/diffusion in reduced zirconia and ceria, we now reexamine the size and charge effect. Although the removal of one lattice oxygen creates a more open pathway, it also leaves the migrating cation less screened, which raises the electrostatic energy. Likewise, while reduction can lower the electrostatic repulsion, it also increases the size of the migrating cation making it more difficult to pass through the crowded saddle point. But if a $V_O^{\cdot\cdot}$ is available for a reduced cation, then both cation size and charge screening are less a problem. Although this explanation seems reasonable, it is countered by the fact that $Y^{3+}$, with a lower valence and a larger size just like $Zr^{3+}$, has a diffusivity comparable to that of $Zr^{4+}$ in $Zr_{1-x}Y_xO_{2-x/2}$. (*27, 33-35*) This is in contrast to $Zr^{3+}$ that diffuses $10^3$-$10^4$ times faster than $Zr^{4+}$. (*12-14*) Since the electronic configuration of $Y^{3+}$ and $Zr^{3+}$ does differ, $4d^0$ for $Y^{3+}$ and $4d^1$ for $Zr^{3+}$, we suspect the synergism may be a quantum-mechanical effect not considered in classical defect physics. Indeed, there is a similar observation in $Ce_{1-x}Gd_xO_{2-x/2}$: absence of Gd/Ce de-mixing under high-temperature electrical loading suggests comparable diffusivity of $Gd^{3+}$ and $Ce^{4+}$, but grain growth study suggests a much higher diffusivity of $Ce^{3+}$ than $Ce^{4+}$. (*13, 16-21*)

We found direct evidence for the quantum effect in the localization of the reduction-generated extra electron. At the ground state, such electron occupies a gap state between the valence band maximum (VBM; mainly O 2*p* orbitals) and the conduction band minimum (CBM; mainly Zr 4*d* orbitals or Ce 4*f* orbitals); it is mostly localized on *d/f* orbital of the target cation (see upper panels of **Fig. 1d-e**; more examples in **Fig. S4a-b, S6a-b, S8a-b** and **S10a-b**). As it comes to the saddle-point state, its energy level relative to the VBM is significantly lowered (see the lower panels of **Fig. 1d-e** with more examples in **Fig. S6d-e, S8d-e** and **S10d-e**). The projected DOS in **Fig. 1d-e** for Model D shows that, from the ground state to the saddle-point state, the Ce/Zr $4d^1/4f^1$ electron lowers its energy by ~1.1/1.7 eV; clearly, such reduction is a major factor in the total migration-barrier lowering of about 1-1.5 eV. This is quite surprising since one would normally expect a reduced cation at a crowed saddle-point with fewer surrounding oxygens and more unsatisfied bonding would have seen more electron-cloud overlap and stronger electron-electron repulsion, which increases the electron energy. It may be viewed as a kind of electronic trebuchet effect, in that the extra electron experiencing a decrease in energy helps a cation migrate much like a counterweight experiencing a decrease in potential energy helps a projectile launch. In the following, we shall examine Ti migration in cubic $BaTiO_3$ to see whether the electronic trebuchet effect for the transitional state is unique to the fluorite structure or not.

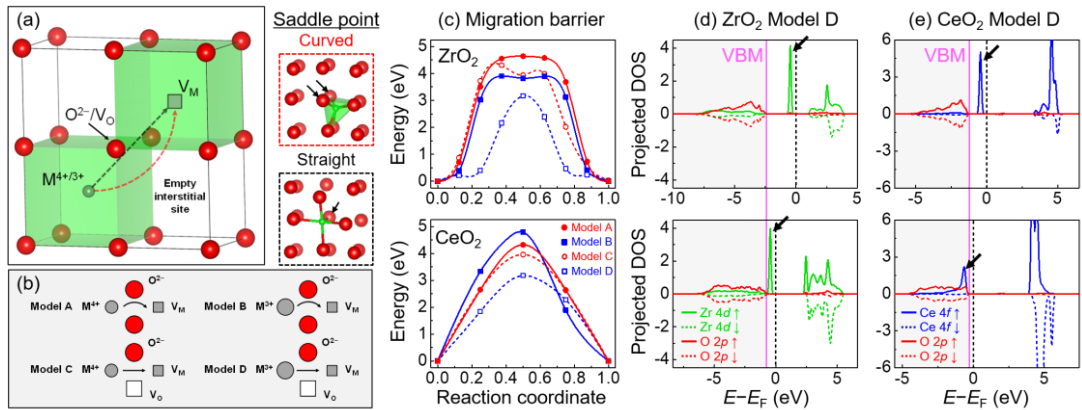

**Figure 1** (a) Schematics of 1×1×½ subcell of cubic $MO_2$ ($M$=Zr and Ce) supercell with one cation vacancy $V_M$. (Produced by VESTA (*31*).) $M^{4+}/M^{3+}$ exchanges location with $V_M$ in straight path (black dashed line) when assisted by oxygen vacancy $V_O$ (marked by arrow) or in curved path (red dashed line) when there is no $V_O$; their saddle-point configurations shown on right panel. (b) Schematic migration models. **A**: $M^{4+}$-$V_M$ exchange without $V_O$. **B**: $M^{3+}$-$V_M$ exchange without $V_O$. **C**: $M^{4+}$-$V_M$ exchange with $V_O$. **D**: $M^{3+}$-$V_M$ exchange with $V_O$. (c) Cation migration energetics along 110 direction for **Model A-D**. Calculated projected DOS of migrating $M^{3+}$ and nearest O for **Model D** in (d) $ZrO_2$ and (e) $CeO_2$: Upper panel for ground state, lower panel for saddle-point state, localized extra electron indicated by arrow.

**Further support: Ti migration in BaTiO$_3$**

In cubic BaTiO$_3$, $M^{4+}$=Ti$^{4+}$ is the slowest diffusing ion and O$^{2-}$ is the fastest one[29]. It is also known that *A*-site vacancy and oxygen vacancy are the most common defects in perovskites. Therefore, we studied their effects on Ti's exchange with Ti's own vacancy. Two cases were considered. First, for an exchange along the (100) direction, we used a similar set of models for $M^{4/3+}$ as before (see A1-D1 in **Fig. 2a-b**). Next, for an exchange along the (110) direction, we used another set of models for $M^{4/3+}$ and $V_{Ba}''$ (see **Fig. 2c-d**): Model A2, $M^{4+}$ migration without a $V_{Ba}''$; Model B2, $M^{3+}$ migration without a $V_{Ba}''$; Model C2, $M^{4+}$ migration with a $V_{Ba}''$; and Model D2, $M^{3+}$ migration with a $V_{Ba}''$. Other cases of 110 migration with a $V_O^{\bullet\bullet}$ and 100 migration with a $V_{Ba}''$ will not be discussed because $V_O^{\bullet\bullet}/V_{Ba}''$ are far from the migrating Ti, so their effects are likely to be marginal. The calculated trajectories and energies defining the saddle point configurations and energy barriers are summarized in **Fig. 3e**, **Fig. S11-S12** (for saddle point configurations) and **Table S2**.

There are several interesting findings here. First, in most cases Ti does not proceed along a straight line even when there is mirror symmetry and no blocking ion is in the way, e.g., in Model C1 and D1. This reflects a strong tendency for the saddle-point Ti$^{4/3+}$, which is small in size yet high in valence, to bond with O$^{2-}$ nearby. Second, there is a nearly monotonic correlation between a decreasing migration barrier and an increasing saddle-point Ti-Ba distance. Indeed, cation reduction and $V_O^{\bullet\bullet}$ only lower the migration barrier from 8.12 eV in Model A1 to 6.53 eV in D1—still very high values for a compound that melts at 1650 ºC—because their Ti-Ba distances remain rather short at the saddle point. Only in Ti 110 migration with a neighboring $V_{Ba}''$, where the saddle-point Ti faces fewer or no surrounding Ba (see **Fig. S12d-f**), is the migration barrier able to reach its lowest, 2.6 eV in Model D2. Third, as before, there is a strong tendency for the extra electron to localize at the saddle-point Ti compared to the ground-state Ti in Model B1, D1, B2 and D2 (**Fig. 2f-g**, **Fig. S14c-d, S16c-d S18c-d** and **S20c-d**). (The extra electron in Model B1 and D1 is delocalized at the ground state.) This not only helps lessen the electrostatic repulsion but also quantum-mechanically provides the localized electron a much lower energy, close to the VBM or even below the VBM as in Model D1. To summarize, notwithstanding the significant influence of *A*-site vacancy on Ti migration (by lessening Ti-Ba repulsion) *(36)*, very similar effects of size, charge and especially electron localization are manifest in $BaTiO_3$.

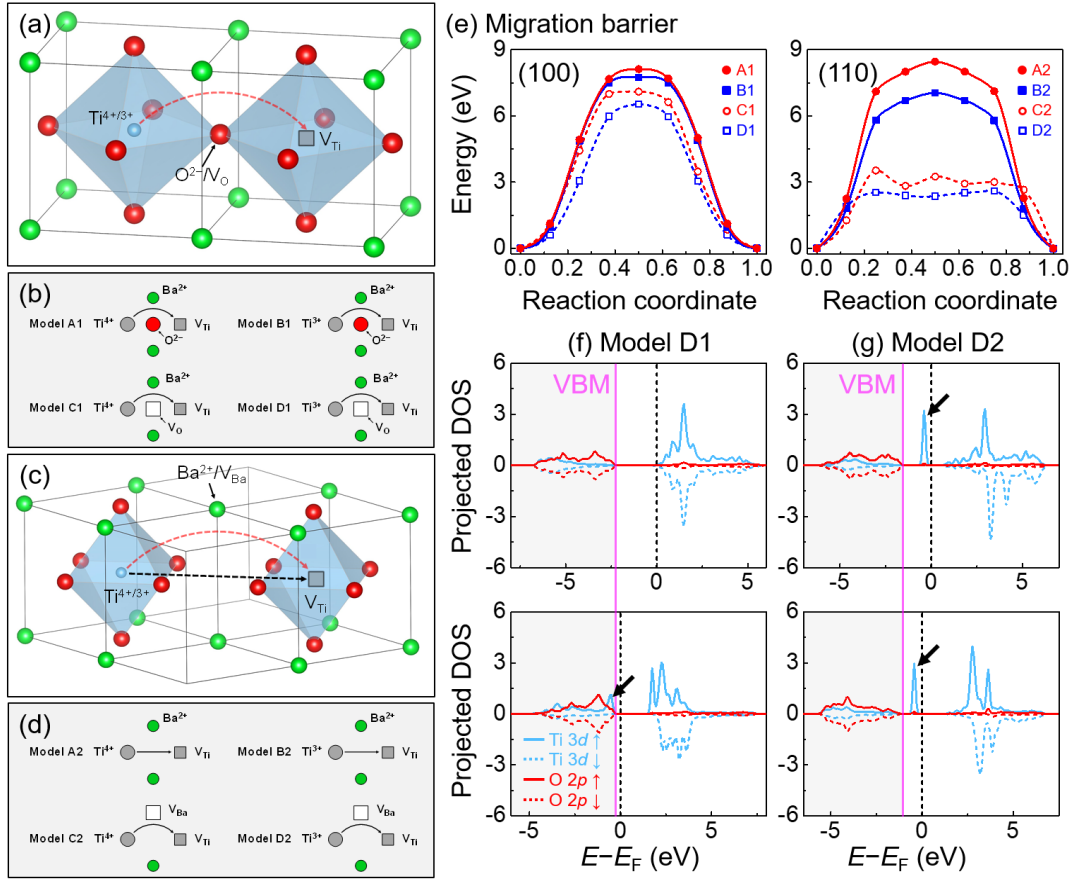

**Figure 2** (a) Schematics of two cubic BaTiO$_3$ cells along 100 direction with one Ti vacancy $V_{Ti}$. Ti$^{4+}$/ Ti$^{3+}$ exchanges location with $V_{Ti}$ in curved path (red dashed line). (b) Schematic migration models. A1: Ti$^{4+}$-$V_{Ti}$ exchange without oxygen vacancy $V_O$. B1: Ti$^{3+}$-$V_{Ti}$ exchange without $V_O$. C1: Ti$^{4+}$-$V_{Ti}$ exchange with $V_O$. D1: Ti$^{3+}$-$V_{Ti}$ exchange with $V_O$. (c) Schematics of two cubic BaTiO$_3$ cells along 110 direction with one $V_{Ti}$. Ti$^{4+}$/ Ti$^{3+}$ exchanges location with $V_{Ti}$ in straight path (black dashed line) when there is no Ba vacancy $V_{Ba}$ (marked by arrow) or in curved path (red dashed line) when assisted by a $V_{Ba}$. (d) Schematic migration models. A2: Ti$^{4+}$-$V_{Ti}$ exchange without $V_{Ba}$. B2: Ti$^{3+}$-$V_{Ti}$ exchange without $V_{Ba}$. C2: Ti$^{4+}$-$V_{Ti}$ exchange with $V_{Ba}$. D2: Ti$^{3+}$-$V_{Ti}$ exchange with $V_{Ba}$. (e) Cation migration energetics along 100 direction for Model A1-D1 and along 110 direction for A2-D2. Calculated projected DOS of migrating $M^{3+}$ and nearest O for (f) Model D1 and (g) D2: Upper panel for **g**round state, lower panel for saddle-point state, localized extra electron indicated by arrow.

**Soft saddle-point state as a negative-$U$ center**

As schematically shown in **Fig. 3a**, our calculation for cation migration depicts an energy landscape flatter in reduced oxides than in unreduced ones. Such a flattening is not just an effect of size and charge, but has a quantum mechanical origin in the energetics of localized electron summarized in **Fig. 3b**. At the ground state, the reduction-generated $d^1/f^1$ electron is either localized at a narrow gap state half-filling the lowest-energy $d/f$ orbital of the migration cation, or delocalized at the CBM, at the energy level $M^L$. In contrast, at the saddle-point state electron localization is always favored, and very importantly its energy level ($M^{SD}<M^L$) is further lowered toward the VBM—sometimes even below the VBM as in Model D in CeO$_2$ and Model D1 in BaTiO$_3$. Loading such an extra electron to the migrating cation as if in a trebuchet provides an energy advantage of $M^L-M^{SD}$ ranging from 0.5 eV to 2.2 eV, which makes it a dominant factor in the total lowering of the migration barrier (about 1 eV) calculated from the total energy of the system.

This quantum effect may be understood in terms of symmetry breaking and lattice relaxation. (i) With a neighboring oxygen or $A$-site vacancy, the saddle-point cation is always asymmetric whereas the ground-state cation at the center of a cube (in fluorite) or an octahedron (in perovskite) has much higher symmetry. Thus, the degenerate $e_g$ ($dx^2\text{-}y^2$, $dz^2$) and $t_{2g}$ ($dxy$, $dxz$, $dyz$) orbitals of the ground-state cation may be further split for the saddle-point cation with the lowest level becoming lower in energy, for example as in **Fig. 3c**. (ii) A reduced cation that acquires a $d^1/f^1$ configuration, or more broadly, a Jahn-Teller configuration, can best take advantage of the above symmetry/energy lowering. Indeed, for this reason such cation tends to spontaneously distort to lift the energy degeneracy, but such a tendency is countered by the elastic stiffness of the environment. Since the saddle-point state has a negative elastic modulus in the direction of migration, it offers less or no resistance in some deformation modes. So the combined electronic and atomic configuration of a reduced cation favorable for spontaneous symmetry/energy lowering is the fundamental reason why it has a lower migration barrier.

The above result can also be understood in terms of negative *U*, first discovered in amorphous conducting chalcogenides that are elastically soft. (*30*) Anderson proposed that they have a large *negative* electron-phonon interaction, so lattice relaxation upon the introduction of an extra electron can lower the system energy to below the value before adding the electron. While reduced oxides are not at all like amorphous chalcogenides, in Model D for $CeO_2$ and Model D1 for $BaTiO_3$, the energy of the extra electron does lie below the VBM. That is, although the successive electrons in the O-2*p* manifold must occupy a higher energy level, including the last electron that fills the top of the VBM, the extra electron that comes next actually occupies a lower energy level in **Fig. 3b** despite its always positive interaction (namely Hubbard *U*) with all the existing electrons already in the manifold. Such energy decrement signals a negative *U*, and it is unique to the saddle point that has an extremely soft environment and experiences a large asymmetric distortion shown in **Fig. 1e** and **2f**. In contrast, a cation in the much stiffer ground state apparently has a positive $U$ in that its extra electron always goes up in energy by an amount of the band gap to lie at the bottom of the CBM.

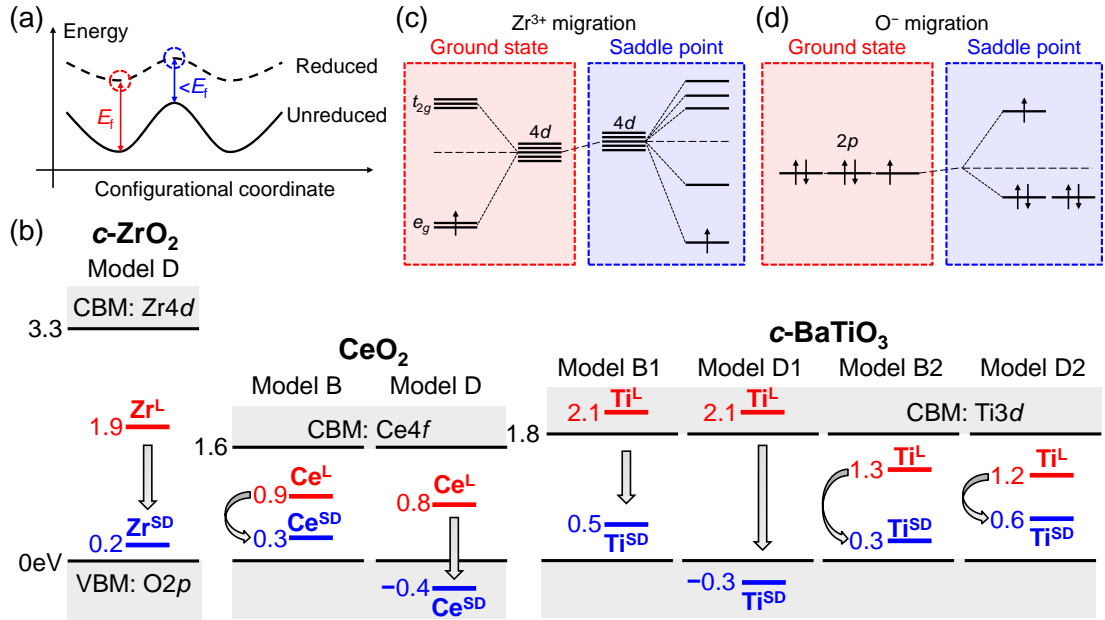

**Figure 3** (a) Schematic energy landscape of reduced and unreduced oxides, from ground state to saddle-point state for cation migration. (b) Energy levels in $ZrO_2$, $CeO_2$ and $BaTiO_3$ with zero energy set at VBM. The level of an extra electron localized at lattice cation in ground state is labeled as $M^L$ ($M$=Zr, Ce, or Ti); the level at migrating cation in saddle-point state is labeled as $M^{SD}$. Counterweight $d/f$ electron provides an energy advantage of ($M^L - M^{SD}$) for cation migration. Schematic crystal field splitting of (c) $4d$ for migrating $Zr^{3+}$ and (d) $2p$ orbitals for migrating $O^-$ in ground state and saddle-point state.

**Estimates and observations of enhancement factor in $ZrO_2$ and $CeO_2$**

In undoped and acceptor-doped zirconia and ceria, the concentration of cation vacancies is extremely low but $V_O^{\bullet\bullet}$ are abundant and fast-diffusing with a migration barrier of only ~0.5 eV. So Model C should prevail under an oxidizing condition, and Model D should prevail in a reducing atmosphere. Without reduction, the cation diffusivity is proportional to $[M^{4+}][V_M]\exp\left(-\dfrac{\text{Barrier of }M^{4+}}{k_B T}\right)$ with $[M^{4+}]$ set to be 1; with reduction, it is proportional to $[M^{3+}][V_M]\exp\left(-\dfrac{\text{Barrier of }M^{3+}}{k_B T}\right)$ with $[M^{3+}]$ depending on doping or the atmosphere. Therefore, the enhancement factor due to reduction is $[M^{3+}]/[M^{4+}]\exp\left(\dfrac{\Delta E}{k_B T}\right)$, where $\Delta E$ is the difference in the barrier height in Model C and D, i.e. 1.14/0.68 eV for $ZrO_2/CeO_2$. When reduction is severe, $[M^{3+}]/[M^{4+}]$ approaches 1, which gives an enhancement factor of 7950 for zirconia and 210 for ceria at 1200 °C. This condition is experimentally met (i) for $Zr_{0.852}Y_{0.148}O_{1.926}$ and $Zr_{0.942}Y_{0.058}O_{1.971}$ under electro-reduction at 1200 °C, whose grain boundary mobility is >1,000 times faster than that in air *(12, 13)*, and (ii) for $Ce_{0.9}Gd_{0.1}O_{1.95}$ in 5% $H_2$ at 1200 °C, whose mobility is 400 times faster than that in air *(13)*. So there is a reasonable agreement between the estimates and the experimental observations. On the other hand, the enhancement factor should be small when $[M^{3+}]/[M^{4+}] \ll 1$. In such a case, $[M^{3+}]/[M^{4+}]$ can be estimated from the thermodynamic data of the reaction $O_O^\times = 2e + V_O^{\bullet\bullet} + \dfrac{1}{2}O_2$ by fixing the $V_O^{\bullet\bullet}$ concentration set by the acceptor dopants *(37, 38)*. It provides $[M^{3+}]/[M^{4+}]$ of $1.7\times10^{-4}$ for $Zr_{0.852}Y_{0.148}O_{1.926}/Zr_{0.942}Y_{0.058}O_{1.971}$ at 1200 °C in an oxygen partial pressure of $10^{-10}$ atm (which is representative for 5% $H_2$); small enhancement factors of 1.5-2 were indeed observed for their grain boundary mobilities in 5% $H_2$ over those in air *(13)*. Likewise, a small $[M^{3+}]/[M^{4+}] \sim 10^{-3}$ at 1300 ºC is seen in $CeO_2$ in air, which explains why its grain boundary mobility in air is only 2 times faster than that in oxygen at the same temperature *(15)*. Lastly, it is known that except for $[M^{3+}]/[M^{4+}]$ approaching 1, the enhancement is always larger in ceria than in zirconia under the same set of atmospheres *(13, 15-21)*, because ceria is much easier to reduce than zirconia.

The above discussion demonstrates that forming $M^{3+}$ can enhance cation diffusion and such an effect is especially potent for $Ce^{3+}$. Furthermore, according to **Fig. 3b** it is also energetically possible for the ground-state $Ce^{3+}$ to donate the extra electron to a saddle-point $Zr^{3+}$. In this way, the more readily formed $Ce^{3+}$ may behave like a catalyst sharing its extra electron with other cations to facilitate their migration. Enhanced mass transport in Ce-containing alloy oxides, which must involve not only $Ce^{3/4+}$ diffusion but also diffusion of all other cations, can thus be understood. The combined reduction-and-catalysis mechanism can also explain the following compositional effects. (i) When 1 mol% $DO_x$ is added to $Zr_{0.88}Ce_{0.12}O_2$ with $D$ varying from 2+ (Ca, Mg) to 3+ (Y, Yb, In, Sc) to 4+ (Ce, Ti) to 5+ (Ta, Nb), the grain boundary mobility in air monotonically increases in the same order because a higher valence $D$ ion promotes Zr/Ce reduction. *(39)* (ii) Although more solute usually causes more solute drag on grain growth, $Zr_{0.88}Ce_{0.12}O_2$ actually has a larger grain size (~2 μm) than $Zr_{0.96}Y_{0.04}O_{1.98}$ (~0.2 μm) because Ce is a diffusion-enhancing catalytic solute and Y is not. (*40*) (iii) In $Al_2O_3$ containing 2-10 vol% $ZrO_2$ as grain-boundary-pinning particles, Ce doping causes rapid coarsening of $ZrO_2$ particles in $H_2$ but not in air, thus triggering a corresponding microstructure transition from large $Al_2O_3$ grains with intragranular $ZrO_2$ in $H_2$ to small $Al_2O_3$ grains with intergranular $ZrO_2$ particles in air. (*40*) (iv) Ceria-based ceramics (e.g., doped $CeO_2$ and $BaCeO_3$-based perovskites) can be sintered at much lower temperatures than zirconia-based ones (e.g., doped $ZrO_2$ and $BaZrO_3$-based perovskites). (*3, 15*) (v) When 0.3 mol% of Cu, Mn, Fe, Co, Cr or Zn is presented in the form of their respective oxides, they all enhance sintering and superplastic flow of $Zr_{0.96}Y_{0.04}O_{1.98}$ because these cations have multiple valences and can easily donate/receive electrons, thus acting as catalytic promoters of Zr/Y migration. (*41, 42*) This may also explain the so-called "activated sintering" phenomenon in which large sintering enhancement is seen at temperatures below the eutectic (melting) temperature of the host oxide and the dopant oxide.

**Other related technologically important observations**

Finally, we point to several technologically important observations that have gone unexplained until now.

(1) Field-assisted sintering: It refers to enhanced sintering under a large slow-varying DC voltage and pressure (the so-called spark plasma sintering) (*43*), a DC voltage and a constant heating rate (the so-called flash sintering) (*5, 44*), or in a microwave heater (*45*). These fields can cause redox reactions in the sample, thus reduction, which enhances diffusion. Moreover, if the field creates any excess electron population but the material is otherwise still near the thermodynamic equilibrium, then cation that captures an extra electron may diffuse faster.

(2) Valence-change (resistance) memory: Resistance memory relying on the formation, rupture and reconnection of conducting filaments has been built using thin films of $Ta_2O_5$, $HfO_2$ and $TiO_2$, among others. Although $O^{2-}$ is usually the fast diffusing species in these oxides, unexpectedly fast cation diffusion when the device is nominally at room temperature has also been reported. (*10*) Since $Ta^{5+}$, $Hf^{4+}$ and $Ti^{4+}$ all have the $d^0$ configuration, this is likely due to their reduced cations that also exist in the reduction filaments.

(3) Suboxide at $Zr/ZrO_2$ interface: Offering diffusional accommodation of the oxidation strain/stress, the suboxide layer can maintain the integrity of the oxide scale and is crucial for the oxidation resistance of Zr metal. (*46, 47*) This is possible because $Zr^{3+}$ cation can diffuse much faster than $Zr^{4+}$. Therefore, while $ZrO_2$ is brittle and tends to crack, diffusional creep/plasticity makes $ZrO_{2-x}$ a good adhesive layer.

(4) Oxygen gas evolution in Li-ion battery: Li-rich cathodes (e.g., $Li_{1.2}Ni_{0.13}Co_{0.13}Mn_{0.54}O_2$) relies on oxygen ion redox ($O^{2-}/O^-$) to deliver high capacities, yet it suffers from continuous oxygen loss and oxygen gas evolution during high-voltage charging/cycling. (*48, 49*) This continuous loss must require oxygen ion diffusion from the bulk to the surface, which is not expected at room temperature because in rock-salt structure oxides anion, not cations, is the slowest diffusing ion. However, we have found in analogy to degeneracy lifting of the lowest-energy *d/f* orbitals of a $d^1/f^1$ cation at the saddle point, there is degeneracy lifting of the highest-energy $2p$ orbitals of a $2p^5$ oxygen (i.e. $O^-$) at the saddle point (**Fig. 3d**), which substantially lowers the total energy of electron-filled orbitals thus offering (hole-localized) $O^-$ a higher mobility than $O^{2-}$. Note that in this case $O^-$ is actually smaller than $O^{2-}$, so both the size effect and the charge effect work in concert to favor the diffusion of a hole-localized anion in an oxidizing atmosphere. (Details of this study will be published elsewhere.)

**Conclusions**

To summarize, we have shown that the synergetic enhancement of cation mobility from lattice vacancy (oxygen or *A*-site cation) and cation reduction is largely due to extra-electron localization and energy-level lowering at the saddle point cation, as if there is an electronic trebuchet effect to aid the transitional state. For cations of $d^0$ or $f^0$ electronic configurations, this is also akin to a negative *U* effect that benefits from asymmetric and soft saddle-point configuration. This finding can be generalized to anions, which may experience enhanced diffusion with the aid of extra-hole localization under oxidizing conditions. This picture of enhanced kinetics mediated by radical-like-ion migration may explain many dynamic observations in oxides and other ceramics under common and extreme (e.g., irradiation during in situ transmission electron microscopy and in nuclear reactors) operating conditions.

**Methods**

All first-principles calculations are based on density functional theory (DFT) using the projector augmented-wave (PAW) method within the Perdew-Burke-Ernzerhof (PBE) generalized gradient approximation (GGA), implemented in the Vienna *ab initio* simulation package (VASP). (50-52) The PAW potentials include the following electrons: $5s^24d^2$ for Zr, $5s^25p^64f^15d^16s^2$ for Ce, $5s^25p^66s^2$ for Ba, $3s^23p^63d^24s^2$ for Ti and $2s^22p^4$ for O. A plane-wave cutoff energy of 400 eV was used and the Brillouin zone was sampled using the Monhorst-Pack scheme with a 3×3×3 *k*-point mesh. The DFT+*U* approach by Dudarev *et al.* (53) was used to describe the energy of localized 4*d* electrons of Zr 4*f* electrons of Ce, and 3*d* electrons for Ti. Specifically, we chose the on-site Coulomb interaction parameter *U*, the on-site exchange interaction parameter *J*, and the effective Hubbard parameter $U_{eff}=U-J$ as follows: *U*=5 eV, *J*= 1 eV and $U_{eff}$=4 eV for Zr 4*d* states, *U*=5 eV, *J*=0 eV and $U_{eff}$=5 eV for Ce 4*f* states, and *U*=5 eV, *J*=0.64 eV and $U_{eff}$=4.36 eV for Ti 3*d* states. (Note that Dudarev *et al.* did not treat *U* and *J* independently, only their difference $U_{eff}=U-J$ is meaningful.)

All calculations were performed under periodic boundary conditions. For the ground state, we used a 2×2×2 supercell containing 32 Zr or Ce and 64 O for cubic $ZrO_2$ and $CeO_2$, respectively, and a 3×3×3 supercell containing 27 Ba, 27 Ti and 81 O for cubic $BaTiO_3$. Cation migration was allowed via a vacancy mechanism because the Schottky reaction, having a much lower defect formation energy than cation Frenkel reaction, provides much more cation vacancies than interstitials in $ZrO_2$, Y-doped $ZrO_2$, $CeO_2$ and $BaTiO_3$. To simulate the migration of a reduced cation, we preselected a target cation next to the cation vacancy and promoted electron localization around it at the ground state by first displacing the neighboring oxygen ions outward by 0.1-1.0 Å, then let the system relax to reach convergence (residue atomic forces less than 0.05 eV/Å). Other models containing oxygen vacancy or Ba vacancy were studied by removing one O or Ba atom from the supercell, respectively.

To track cation migration, the climbing-image nudged-elastic-band (NEB) method implemented in VASP was used with a fixed supercell size and shape. *(54)* In cubic $ZrO_2$ and $BaTiO_3$, it determined the migration path and the barrier with the path defined by 7 intermediate states in addition to the initial and final configurations; in $CeO_2$, 3 intermediate states were specified. Convergence for NEB calculations was considered achieved when the residual atomic forces are less than 0.1 eV/Å. After the NEB calculation, static calculations were conducted to calculate the density of state (DOS) and projected density of state (pDOS) on each atom for the ground-state and the saddle-point configurations with a 6×6×6 *k*-point mesh.

DFT calculations always specify the Fermi level, which is used to determine electron state occupancy. However, to compare electron states in different structural states, such as ground state and saddle-point state, or states with and without one missing cation or O, it is convenient if we can find a reference energy level that is relatively insensitive to the structures/defects. All the structures studied here comprise of cation polyhedra bounded by oxygen ions that are (edge-) interconnected into a continuous three-dimensional network. Moreover, their valence bands are mainly made of O2*p* orbitals. Therefore, we may regard the valence band manifold as representative of network's electronic states. The overall electronic energy of this network, which is continuous, should be relatively insensitive to isolated structural defects/distortions. In this work, we used the valence band maximum as the reference energy to compare electron energies of different structural states.

**Acknowledgements**

Work by Y.D. was performed in partial fulfillment of the PhD requirements at the University of Pennsylvania for which Y.D. and I.W.C. acknowledge support by the Department of Energy (BES grant no. DEFG02-11ER46814) and the facility (LRSM) support by the U.S. National Science Foundation (grant no. DMR-1120901). Q.L. acknowledges support by the startup funding from University of Michigan. J.L. acknowledges support by the Department of Energy (grant no. DE-SC0002633) and a subcontract of the U.S. Department of Energy, Office of Energy Efficiency and Renewable Energy, Advanced Manufacturing Office (AMO) R&D Projects Emerging Research Exploration, under Department of Energy DOE Idaho Operations Office with contract no. DE-AC07-05ID14517.

# Electron Localization Enhances Cation Diffusion in Transition Metal Oxides: An Electronic Trebuchet Effect

Yanhao Dong[1], Liang Qi[2], Ju Li[1,3], and I-Wei Chen[4*]

[1]Department of Nuclear Science and Engineering, Massachusetts Institute of Technology, Cambridge, MA 02139, USA

[2]Department of Materials Science and Engineering, University of Michigan, Ann Arbor, MI 48109, USA

[3]Department of Materials Science and Engineering, Massachusetts Institute of Technology, Cambridge, MA 02139, USA

[4]Department of Materials Science and Engineering, University of Pennsylvania, Philadelphia, PA 19104, USA

**Table of content**

**Table S1** Results of cation migration in cubic $ZrO_2$ and $CeO_2$ via vacancy mechanism. In our GGA+$U$ calculations of defect-free structures, reference Zr-O bond length is 2.24 Å, and Ce-O is 2.36 Å. With cation vacancy in the ground state, these bond lengths shorten. To calculate Ce's Bader charge, we exclude 8$e$ from $5s^2 5p^6$ inner-shell.

| Material | Model | Migration path | Migration barrier (eV) | Nearest cation-oxygen distance (Å) | | Bader charge on migrating cation ($e$) | |
|---|---|---|---|---|---|---|---|
| | | | | Ground state | Saddle point | Ground state | Saddle point |
| $ZrO_2$ | A | Curved | 4.64 | 2.15 | 1.97 | 0.62 | 0.80 |
| | B | Curved | 3.90 | 2.23 | 2.04 | 1.44 | 1.28 |
| | C | Straight | 4.31 | 2.15 | 2.05 | 0.59 | 0.68 |
| | D | Straight | 3.17 | 2.24 | 2.11 | 1.43 | 1.49 |
| $CeO_2$ | A | Curved | 4.32 | 2.27 | 2.03 | 1.63 | 1.74 |
| | B | Curved | 4.79 | 2.39 | 2.12 | 1.91 | 1.92 |
| | C | Straight | 3.96 | 2.25 | 2.06 | 1.65 | 1.67 |
| | D | Straight | 3.28 | 2.38 | 2.13 | 1.91 | 1.92 |

**Table S2** Results of cation migration in cubic BaTiO$_3$ via a vacancy mechanism. In our GGA+$U$ calculations of defect-free cubic BaTiO$_3$, reference Ti-O bond length is 2.01 Å and Ti-Ba distance is 3.48 Å. With cation vacancy in ground state, Ti-O bond length and Ti-Ba distance in ground state, these lengths differ from reference ones. To calculate Ti's Bader charge, we exclude 8$e$ from $3s^23p^6$ inner-shell. * indicates Ti-Ba distance in intermediate state for Model C2 and D2.

| Migration direction | Model | Migration path | Migration barrier (eV) | Nearest Ti-O distance (Å) | | Nearest Ti-Ba distance (Å) | | Bader charge on migrating cation ($e$) | |
|---|---|---|---|---|---|---|---|---|---|
| | | | | Ground | Saddle | Ground | Saddle | Ground | Saddle |
| 100 | A1 | Curved | 8.12 | 1.82 | 1.68 | 3.46 | 2.77×2 | 1.84 | 1.88 |
| | B1 | Curved | 7.10 | 1.82 | 1.73 | 3.48 | 2.79×2 | 1.84 | 2.18 |
| | C1 | Curved | 7.75 | 1.81 | 1.89 | 3.43 | 2.62 | 1.80 | 1.87 |
| | D1 | Curved | 6.53 | 1.81 | 1.94 | 3.44 | 2.84/2.86 | 1.81 | 2.18 |
| 110 | A2 | Straight | 8.46 | 1.99 | 1.89 | 3.56 | 2.61×2 | 1.68 | 2.03 |
| | B2 | Straight | 7.04 | 2.04 | 1.95 | 3.50 | 2.68×2 | 2.03 | 2.19 |
| | C2 | Curved | 3.53 | 1.98 | 1.78 | 3.49 | 3.44×2 (3.10*) | 1.75 | 1.83 |
| | D2 | Curved | 2.60 | 2.06 | 1.86 | 3.48 | 3.42×2 (3.01*) | 1.96 | 2.11 |

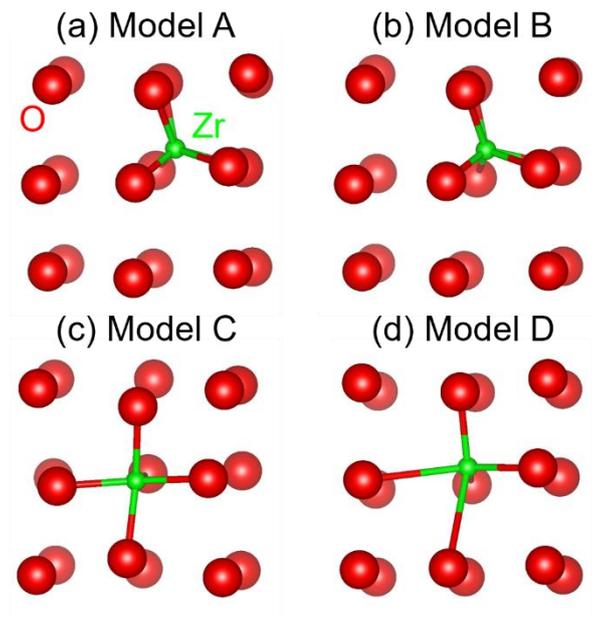

**Figure S1** Atomic arrangements at the saddle point for Zr migration in cubic zirconia of (a) Model A, (b) Model B, (c) Model C and (d) Model D. Zr atom in green, and O atom in red.

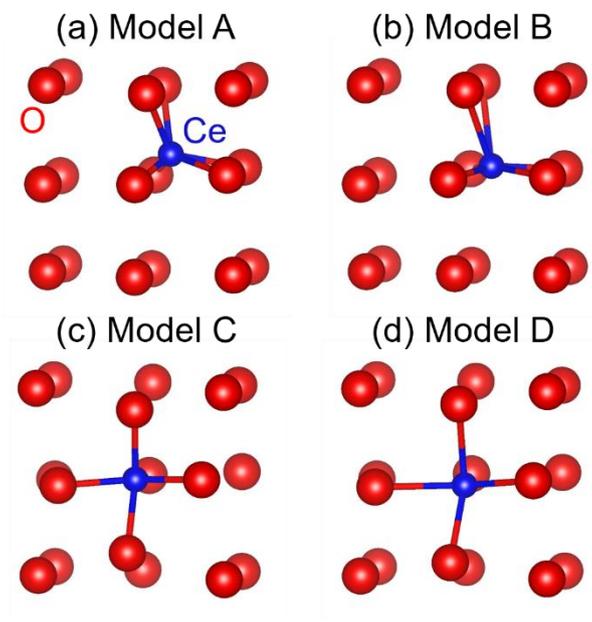

**Figure S2** Atomic arrangements at the saddle point for Ce migration in ceria of (a) Model A, (b) Model B, (c) Model C and (d) Model D. Ce atom in blue, and O atom in red.

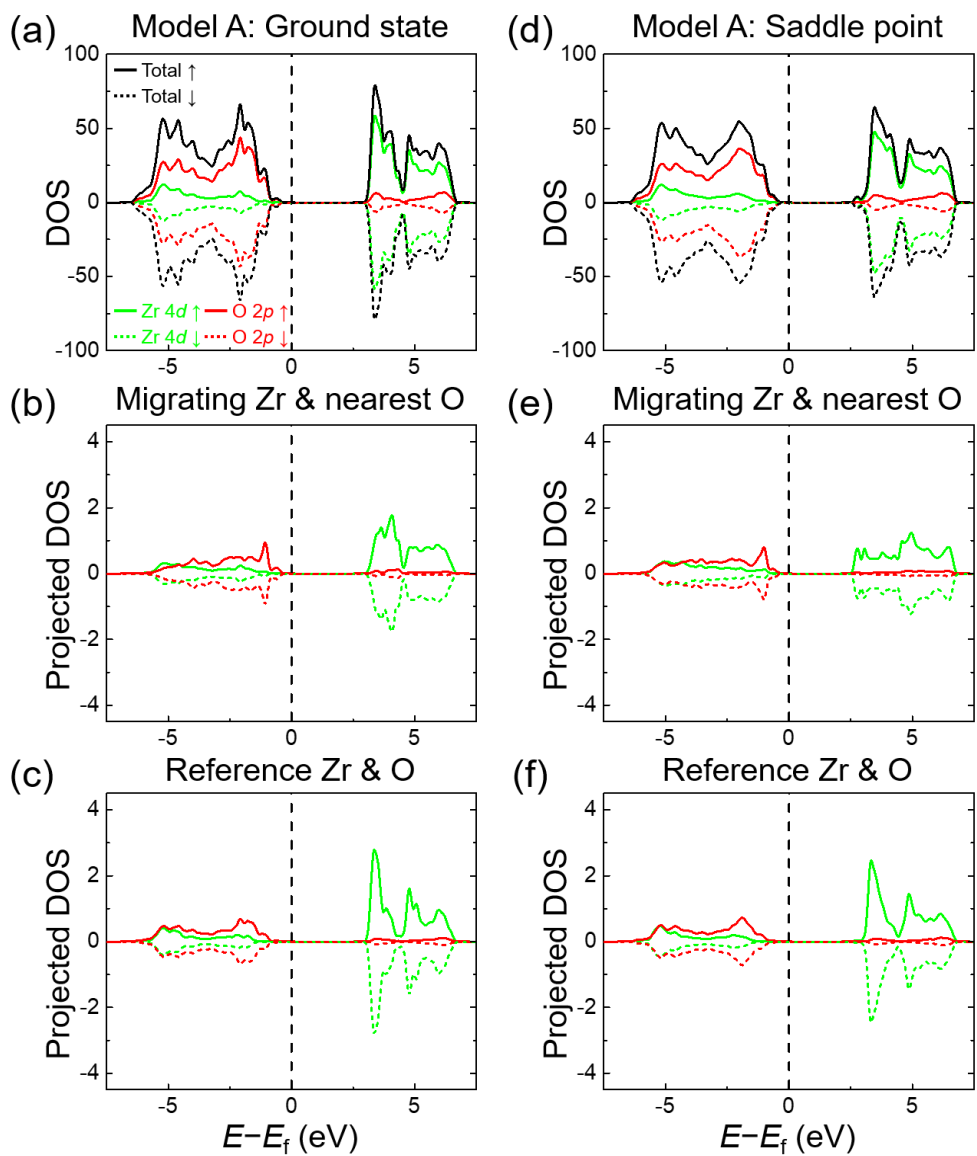

**Figure S3** Calculated density of states (DOS) of cubic zirconia for $Zr^{4+}$ migration **Model A**. Ground state: (a) total DOS, (b) projected DOS of (to be) migrating Zr (in green) and nearest O (in red), and (c) projected DOS of non-participating reference Zr (in green) and O (in red). Saddle-point state: (d) total DOS, (e) projected DOS of migrating Zr (in green) and nearest O (in red), and (f) projected DOS of non-participating reference Zr (in green) and O (in red). In each figure, Fermi energy is set to be zero and spin-up and spin-down states are plotted as positive and negative DOS, respectively.

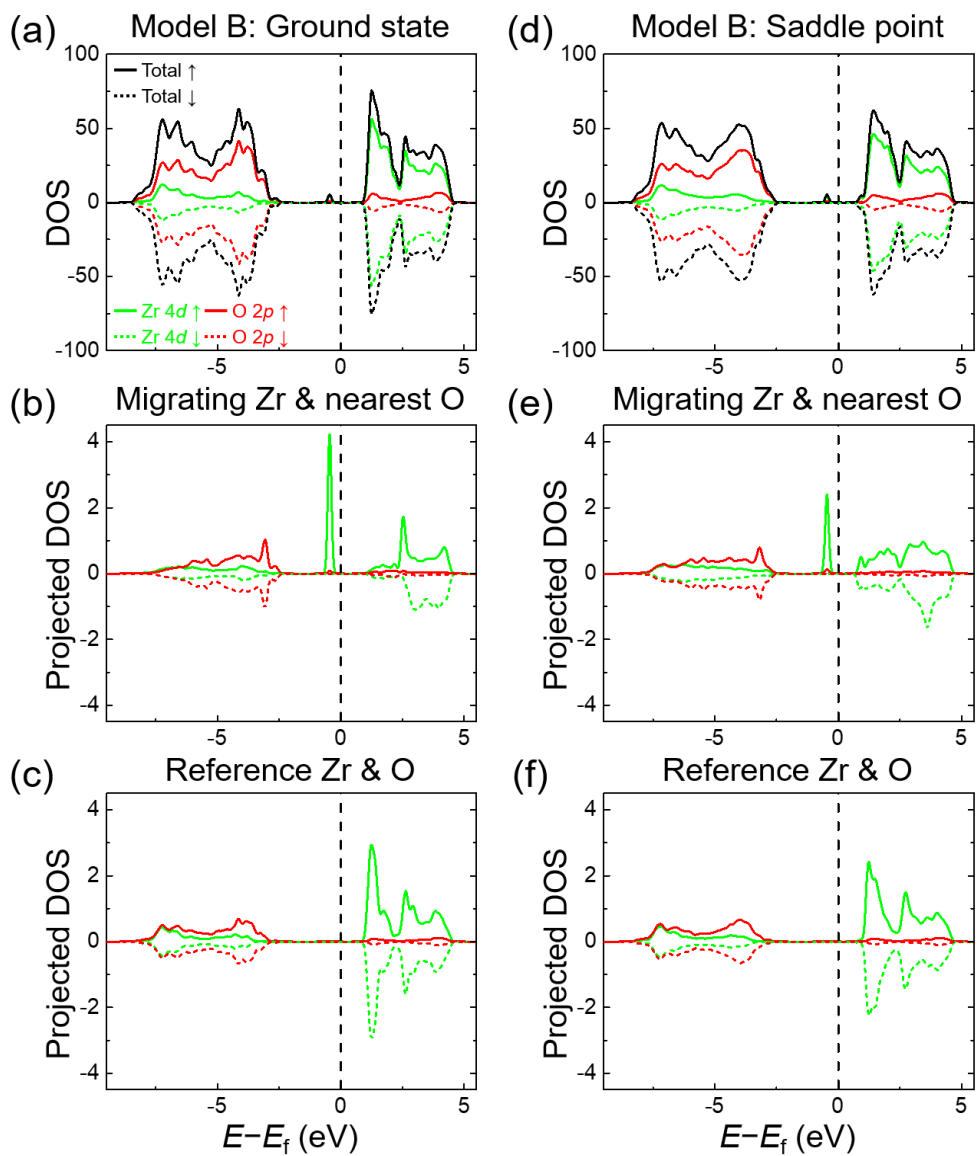

**Figure S4** Calculated density of states (DOS) of cubic zirconia for $Zr^{3+}$ migration **Model B**. Ground state: (a) total DOS, (b) projected DOS of (to be) migrating Zr (in green) and nearest O (in red), and (c) projected DOS of non-participating reference Zr (in green) and O (in red). Saddle-point state: (d) total DOS, (e) projected DOS of migrating Zr (in green) and nearest O (in red), and (f) projected DOS of non-participating reference Zr (in green) and O (in red). In each figure, Fermi energy is set to be zero and spin-up and spin-down states are plotted as positive and negative DOS, respectively.

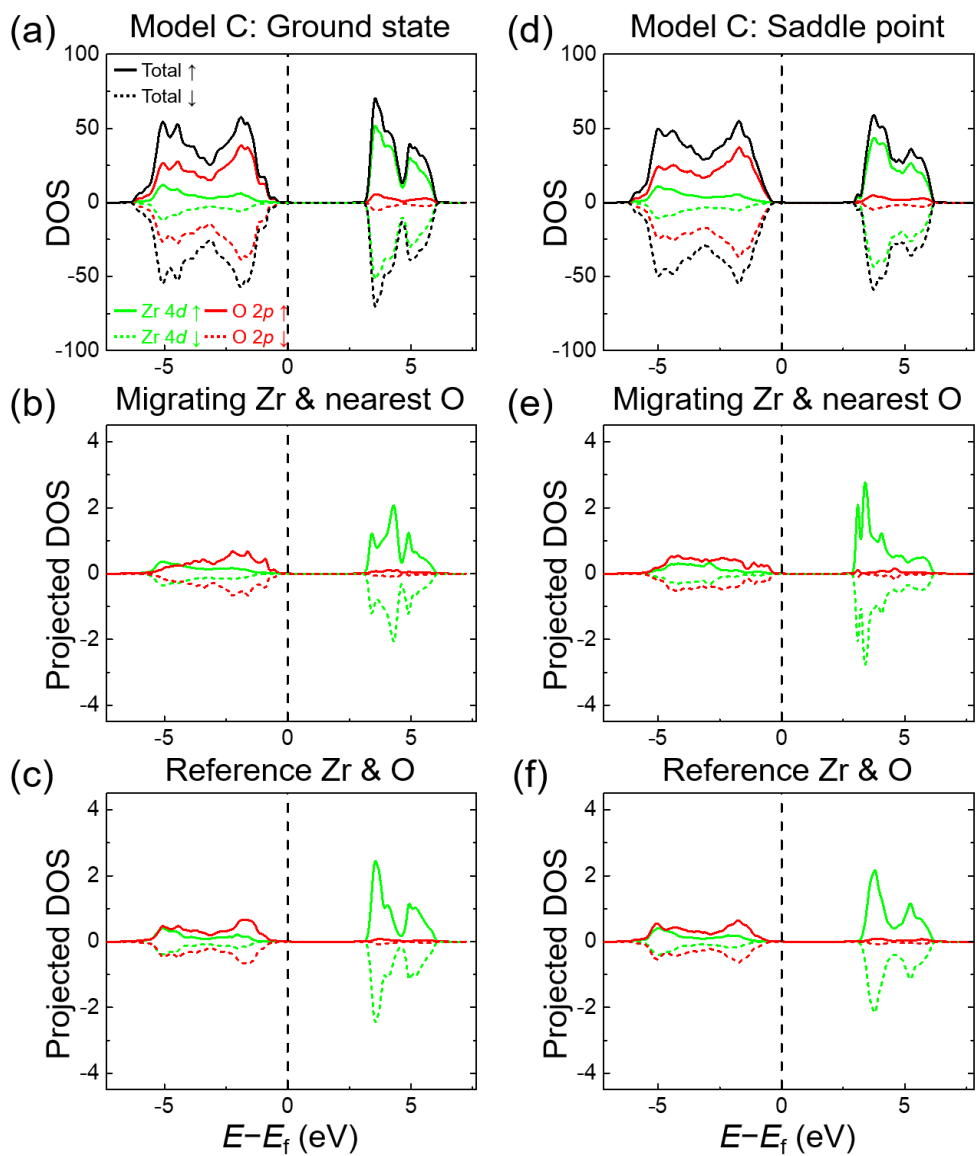

**Figure S5** Calculated density of states (DOS) of cubic zirconia for $Zr^{4+}$ migration **Model C**. Ground state: (a) total DOS, (b) projected DOS of (to be) migrating Zr (in green) and nearest O (in red), and (c) projected DOS of non-participating reference Zr (in green) and O (in red). Saddle-point state: (d) total DOS, (e) projected DOS of migrating Zr (in green) and nearest O (in red), and (f) projected DOS of non-participating reference Zr (in green) and O (in red). In each figure, Fermi energy is set to be zero and spin-up and spin-down states are plotted as positive and negative DOS, respectively.

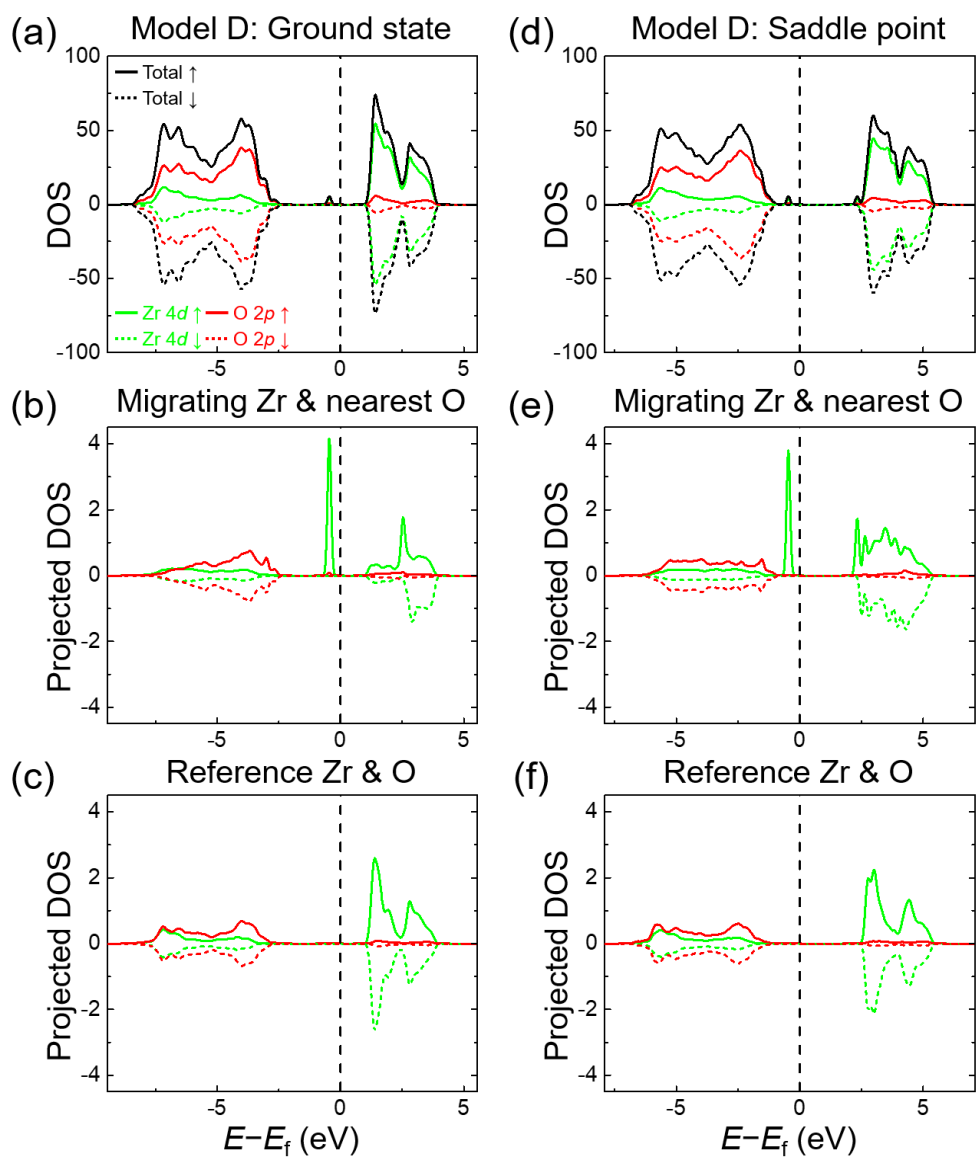

**Figure S6** Calculated density of states (DOS) of cubic zirconia for $Zr^{3+}$ migration **Model D**. Ground state: (a) total DOS, (b) projected DOS of (to be) migrating Zr (in green) and nearest O (in red), and (c) projected DOS of non-participating reference Zr (in green) and O (in red). Saddle-point state: (d) total DOS, (e) projected DOS of migrating Zr (in green) and nearest O (in red), and (f) projected DOS of non-participating reference Zr (in green) and O (in red). In each figure, Fermi energy is set to be zero and spin-up and spin-down states are plotted as positive and negative DOS, respectively.

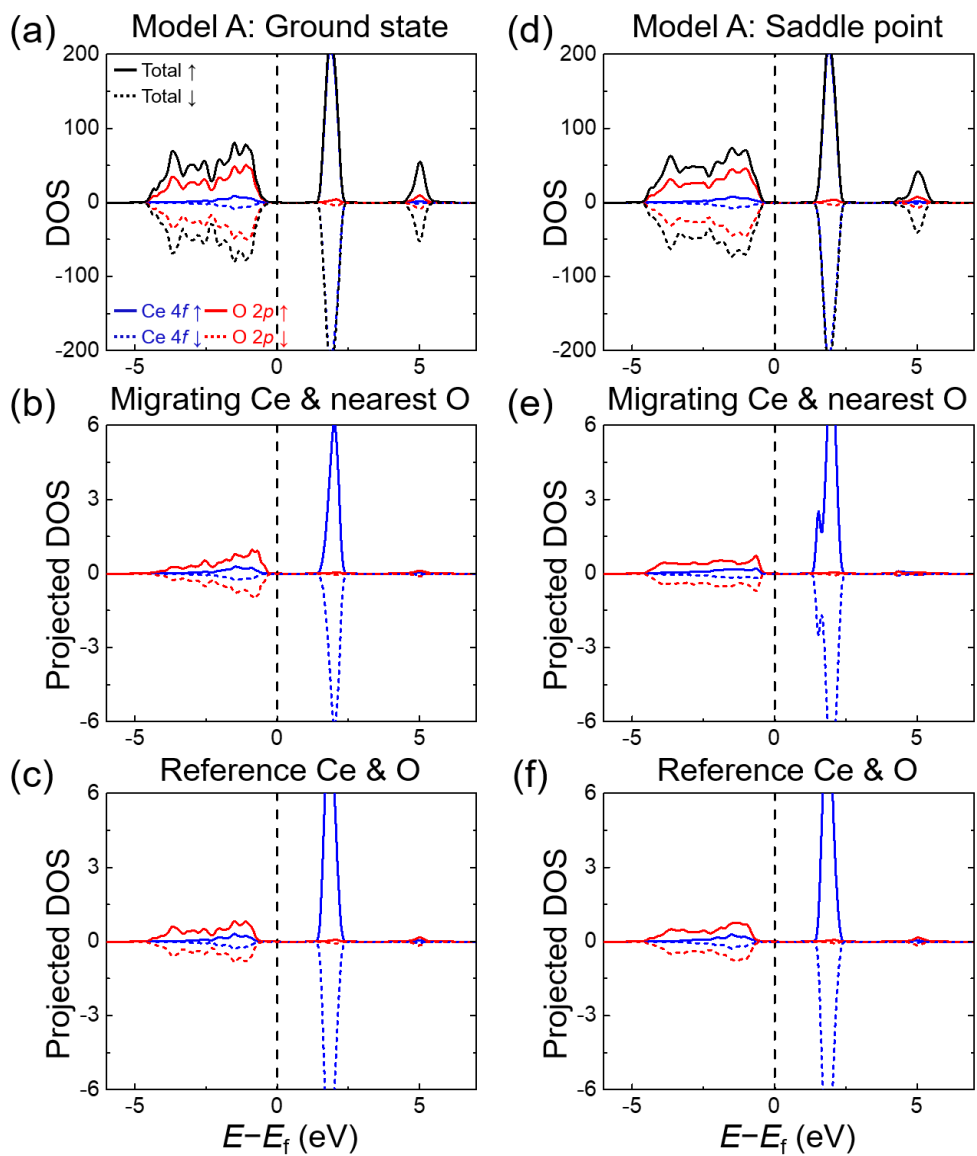

**Figure S7** Calculated density of states (DOS) of ceria for $Ce^{4+}$ migration **Model A**. Ground state: (a) total DOS, (b) projected DOS of (to be) migrating Ce (in blue) and nearest O (in red), and (c) projected DOS of non-participating reference Ce (in blue) and O (in red). Saddle-point state: (d) total DOS, (e) projected DOS of migrating Ce (in blue) and nearest O (in red), and (f) projected DOS of non-participating reference Ce (in blue) and O (in red). In each figure, Fermi energy is set to be zero and spin-up and spin-down states are plotted as positive and negative DOS, respectively.

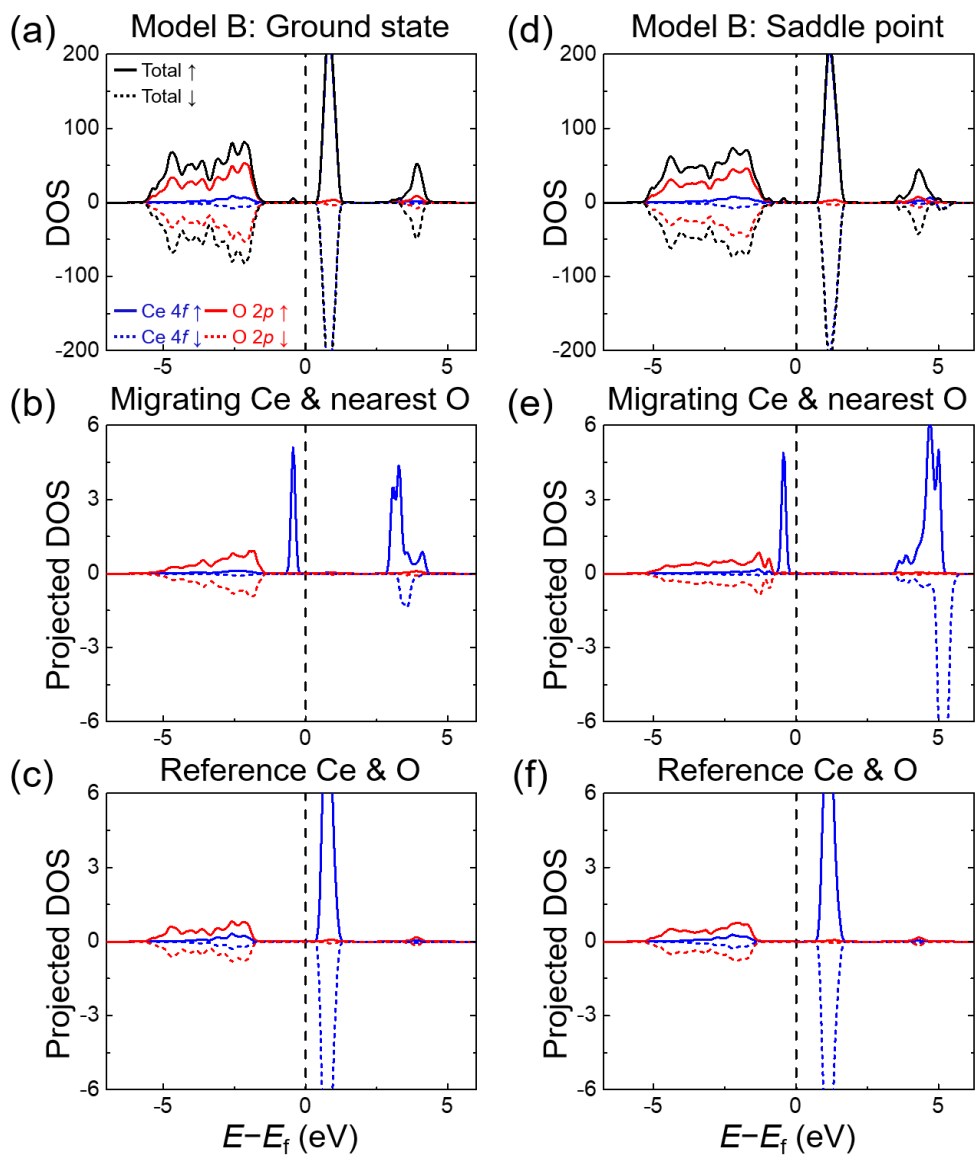

**Figure S8** Calculated density of states (DOS) of ceria for $Ce^{3+}$ migration **Model B**. Ground state: (a) total DOS, (b) projected DOS of (to be) migrating Ce (in blue) and nearest O (in red), and (c) projected DOS of non-participating reference Ce (in blue) and O (in red). Saddle-point state: (d) total DOS, (e) projected DOS of migrating Ce (in blue) and nearest O (in red), and (f) projected DOS of non-participating reference Ce (in blue) and O (in red). In each figure, Fermi energy is set to be zero and spin-up and spin-down states are plotted as positive and negative DOS, respectively.

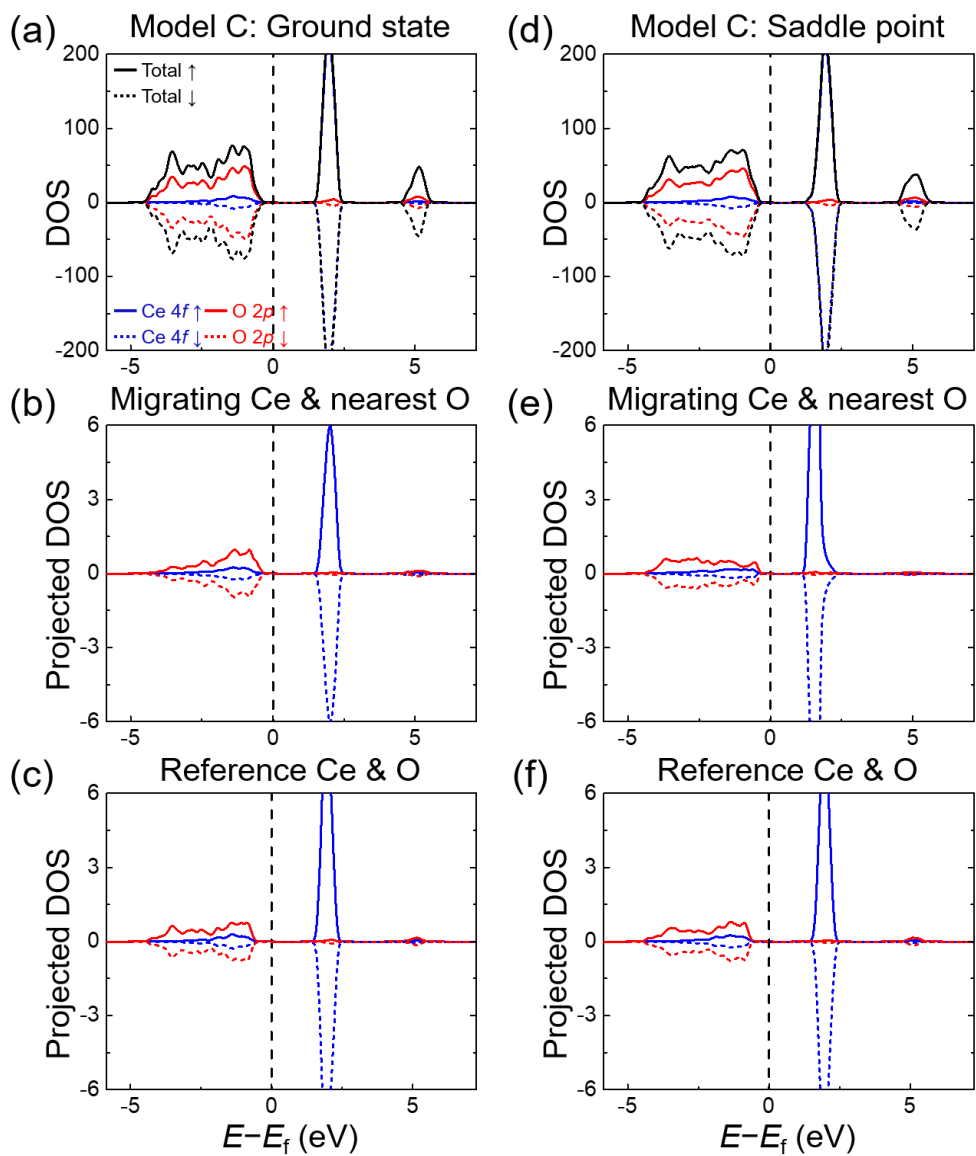

**Figure S9** Calculated density of states (DOS) of ceria for Ce$^{4+}$ migration **Model C**. Ground state: (a) total DOS, (b) projected DOS of (to be) migrating Ce (in blue) and nearest O (in red), and (c) projected DOS of non-participating reference Ce (in blue) and O (in red). Saddle-point state: (d) total DOS, (e) projected DOS of migrating Ce (in blue) and nearest O (in red), and (f) projected DOS of non-participating reference Ce (in blue) and O (in red). In each figure, Fermi energy is set to be zero and spin-up and spin-down states are plotted as positive and negative DOS, respectively.

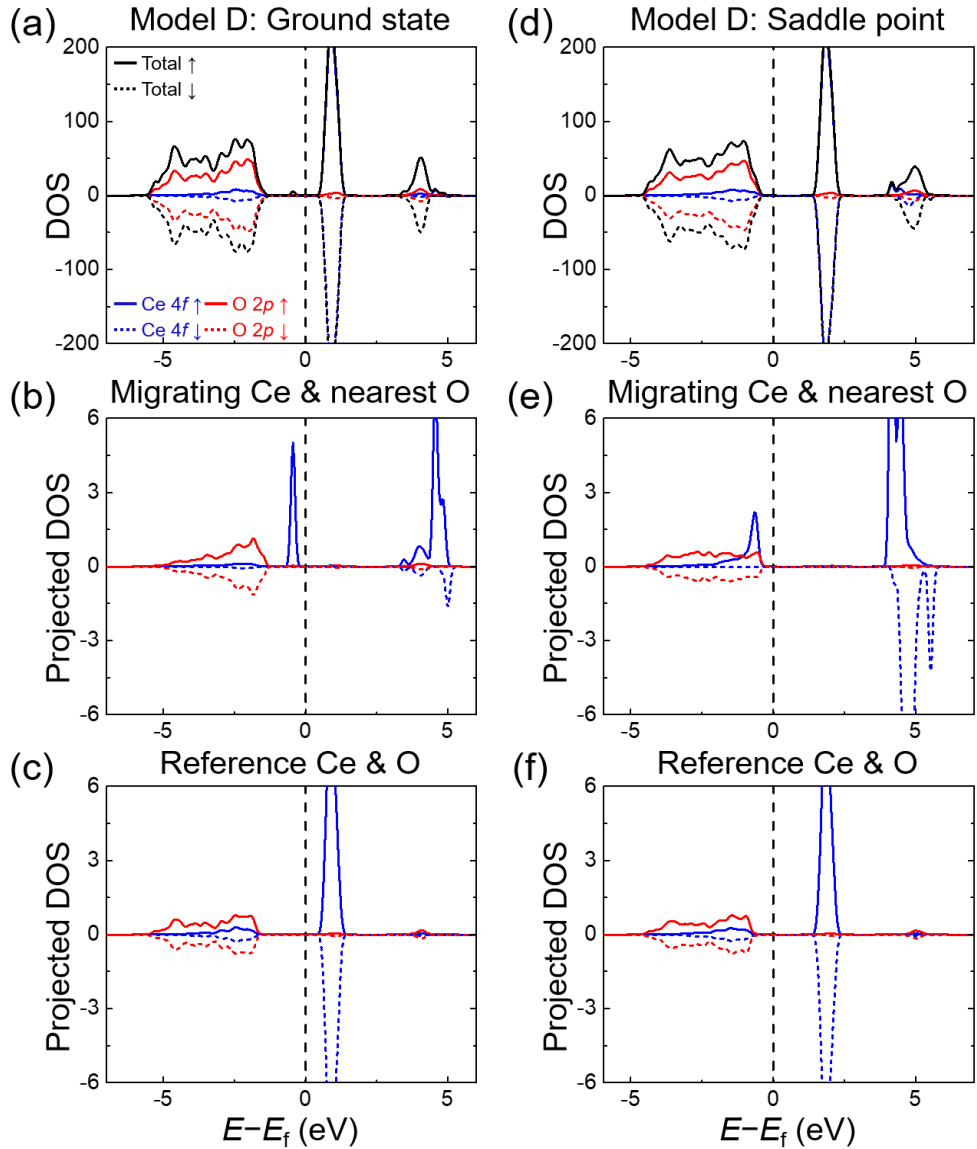

**Figure S10** Calculated density of states (DOS) of ceria for $Ce^{3+}$ migration **Model D**. Ground state: (a) total DOS, (b) projected DOS of (to be) migrating Ce (in blue) and nearest O (in red), and (c) projected DOS of non-participating reference Ce (in blue) and O (in red). Saddle-point state: (d) total DOS, (e) projected DOS of migrating Ce (in blue) and nearest O (in red), and (f) projected DOS of non-participating reference Ce (in blue) and O (in red). In each figure, Fermi energy is set to be zero and spin-up and spin-down states are plotted as positive and negative DOS, respectively.

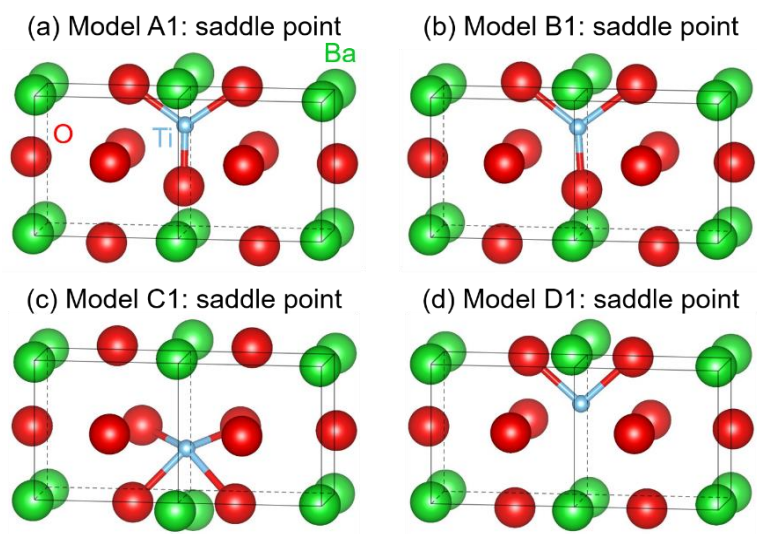

**Figure S11** Atomic arrangements at the saddle point for Ti 100 migration in cubic $BaTiO_3$ of (a) Model A1, (b) Model B1, (c) Model C1 and (d) Model D1. Ti atom in blue, Ba atom in green and O atom in red.

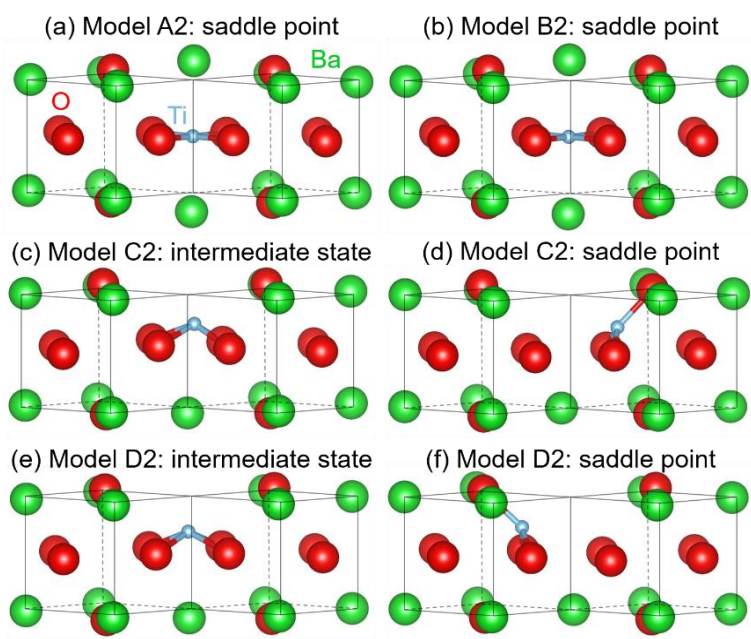

**Figure S12** Atomic arrangements at the saddle point for Ti 110 migration in cubic $BaTiO_3$ of (a) Model A2, (b) Model B2, (d) Model C2 and (f) Model D2. Also shown are atomic arrangement at intermediate state of (c) Model C2 and (e) Model D2, which are structurally similar to saddle point of Model A2 and B2. Ti atom in blue, Ba atom in green and O atom in red.

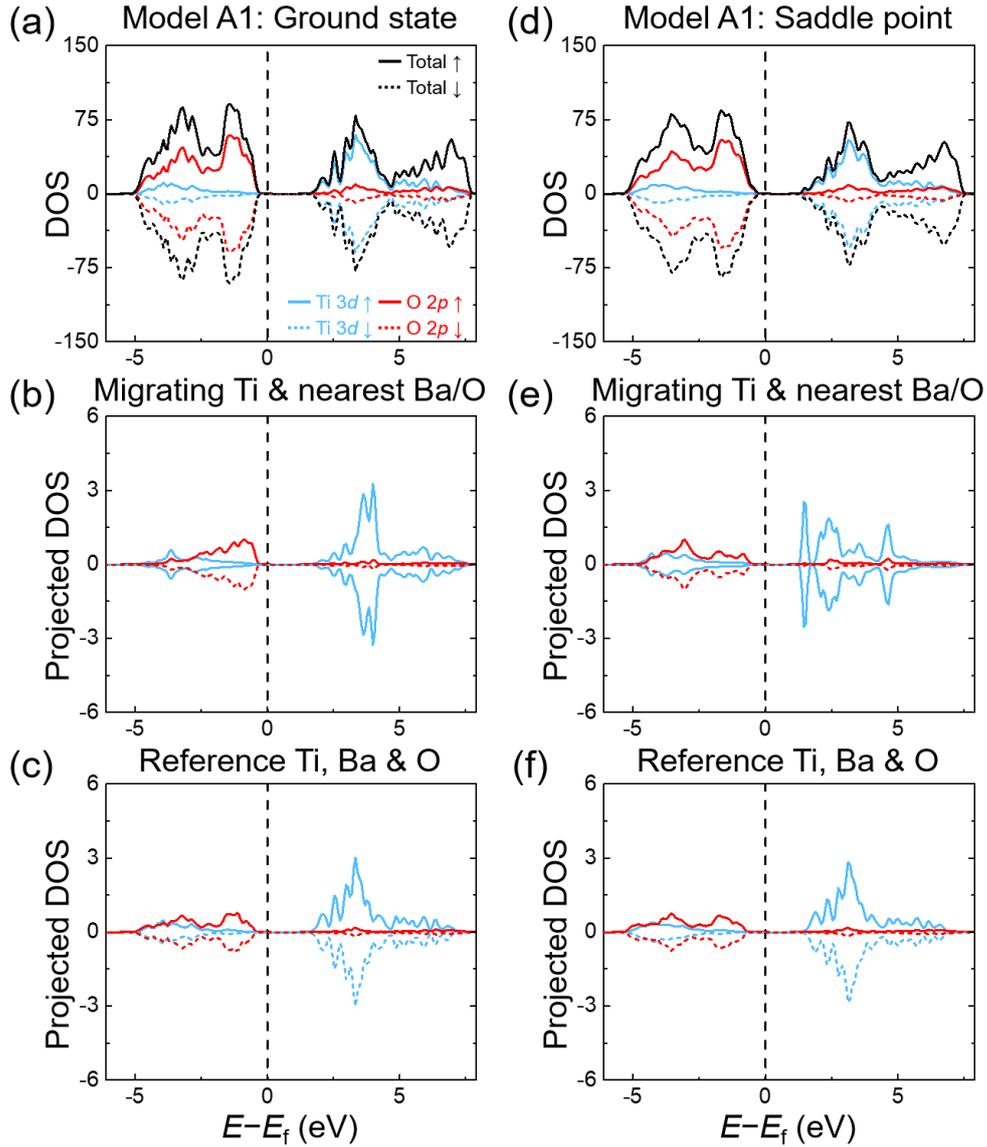

**Figure S13** Calculated density of states (DOS) of cubic BaTiO$_3$ for Ti$^{4+}$ 100 migration **Model A1**. Ground state: (a) total DOS, (b) projected DOS of (to be) migrating Ti (in blue) and nearest O (in red), and (c) projected DOS of non-participating reference Ti (in blue) and O (in red). Saddle-point state: (d) total DOS, (e) projected DOS of migrating Ti (in blue) and nearest O (in red), and (f) projected DOS of non-participating reference Ti (in blue) and O (in red) at the saddle-point configuration. In each figure, Fermi energy is set to be zero and spin-up and spin-down states are plotted as positive and negative, respectively.

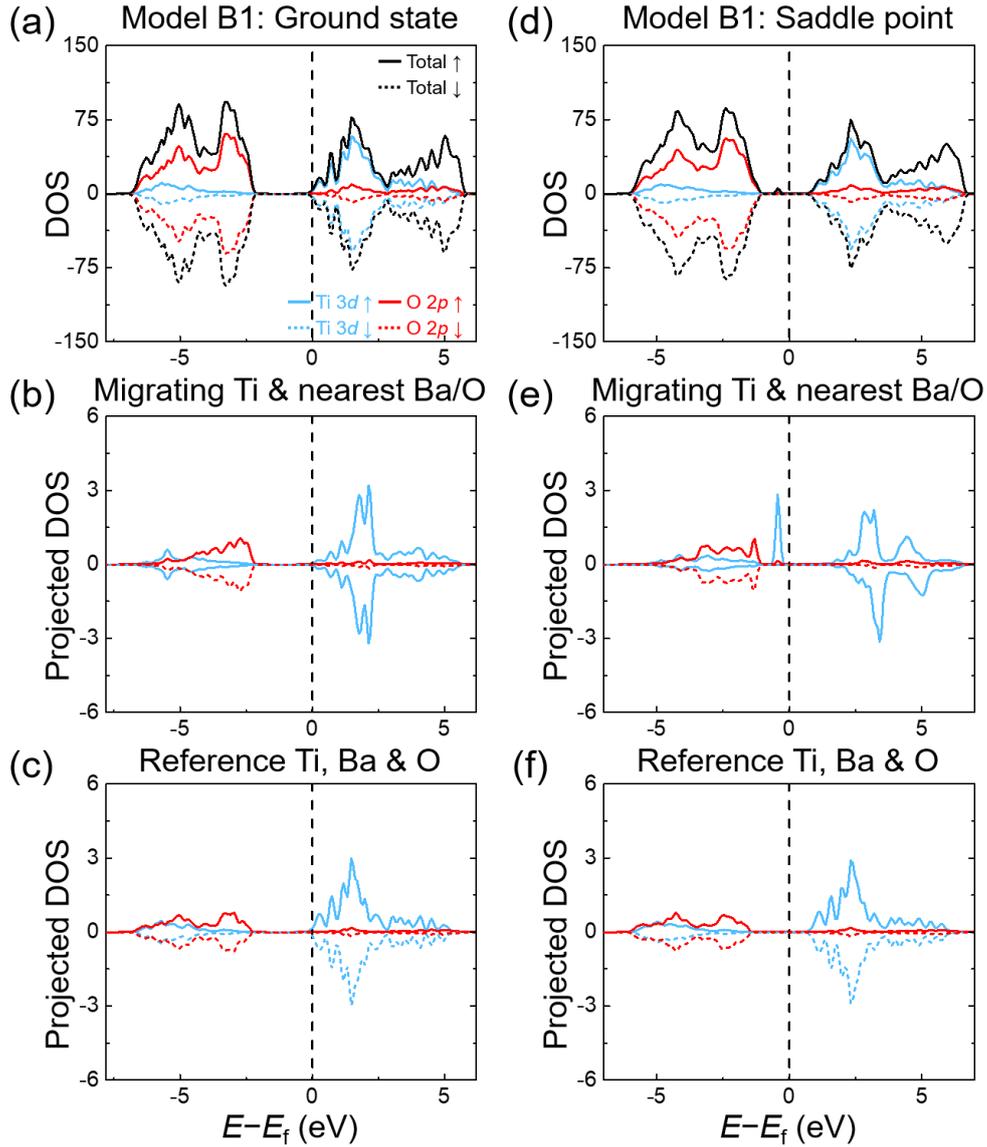

**Figure S14** Calculated density of states (DOS) of cubic BaTiO$_3$ for Ti$^{3+}$ 100 migration **Model B1**. Ground state: (a) total DOS, (b) projected DOS of (to be) migrating Ti (in blue) and nearest O (in red), and (c) projected DOS of non-participating reference Ti (in blue) and O (in red). Saddle-point state: (d) total DOS, (e) projected DOS of migrating Ti (in blue) and nearest O (in red), and (f) projected DOS of non-participating reference Ti (in blue) and O (in red) at the saddle-point configuration. In each figure, Fermi energy is set to be zero and spin-up and spin-down states are plotted as positive and negative, respectively.

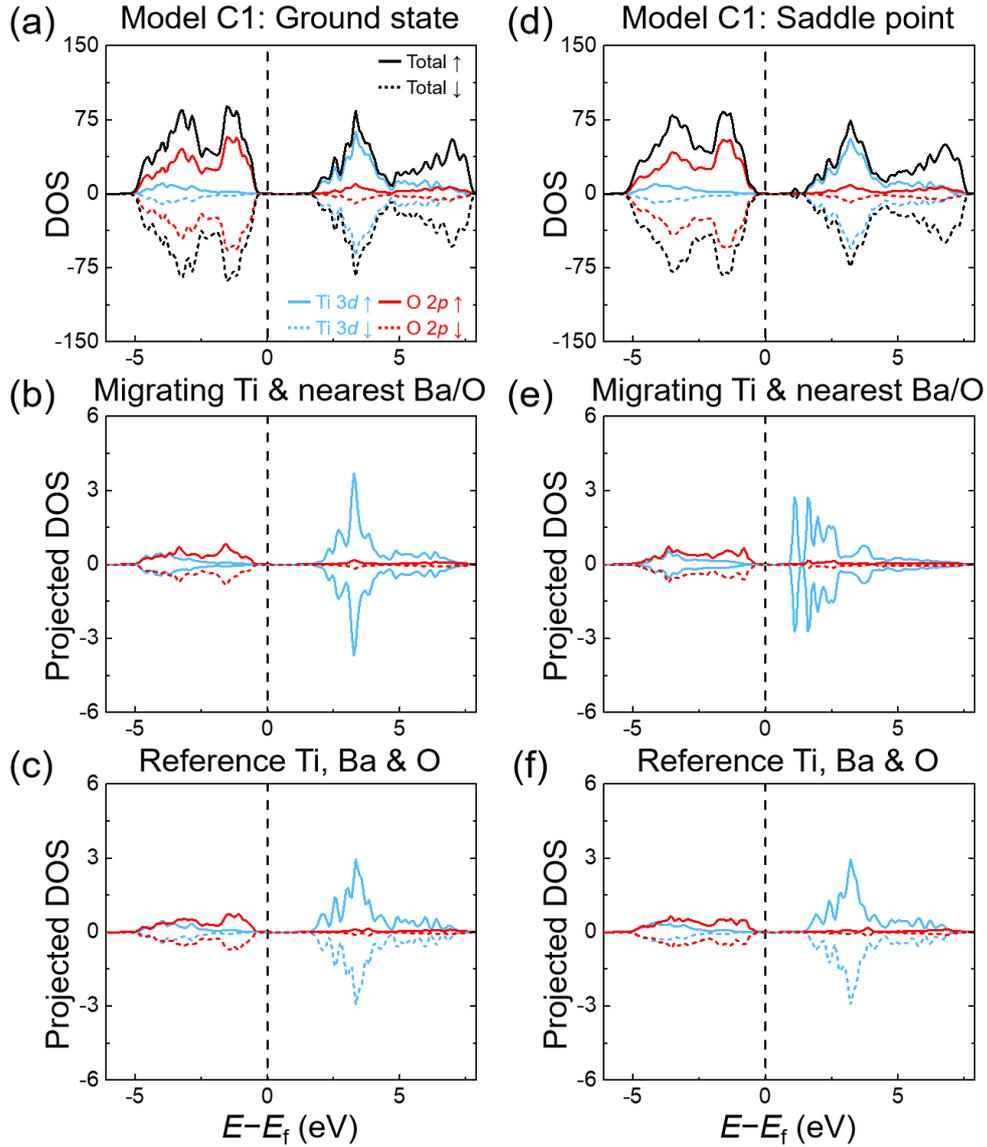

**Figure S15** Calculated density of states (DOS) of cubic BaTiO$_3$ for Ti$^{4+}$ 100 migration **Model C1**. Ground state: (a) total DOS, (b) projected DOS of (to be) migrating Ti (in blue) and nearest O (in red), and (c) projected DOS of non-participating reference Ti (in blue) and O (in red). Saddle-point state: (d) total DOS, (e) projected DOS of migrating Ti (in blue) and nearest O (in red), and (f) projected DOS of non-participating reference Ti (in blue) and O (in red) at the saddle-point configuration. In each figure, Fermi energy is set to be zero and spin-up and spin-down states are plotted as positive and negative, respectively.

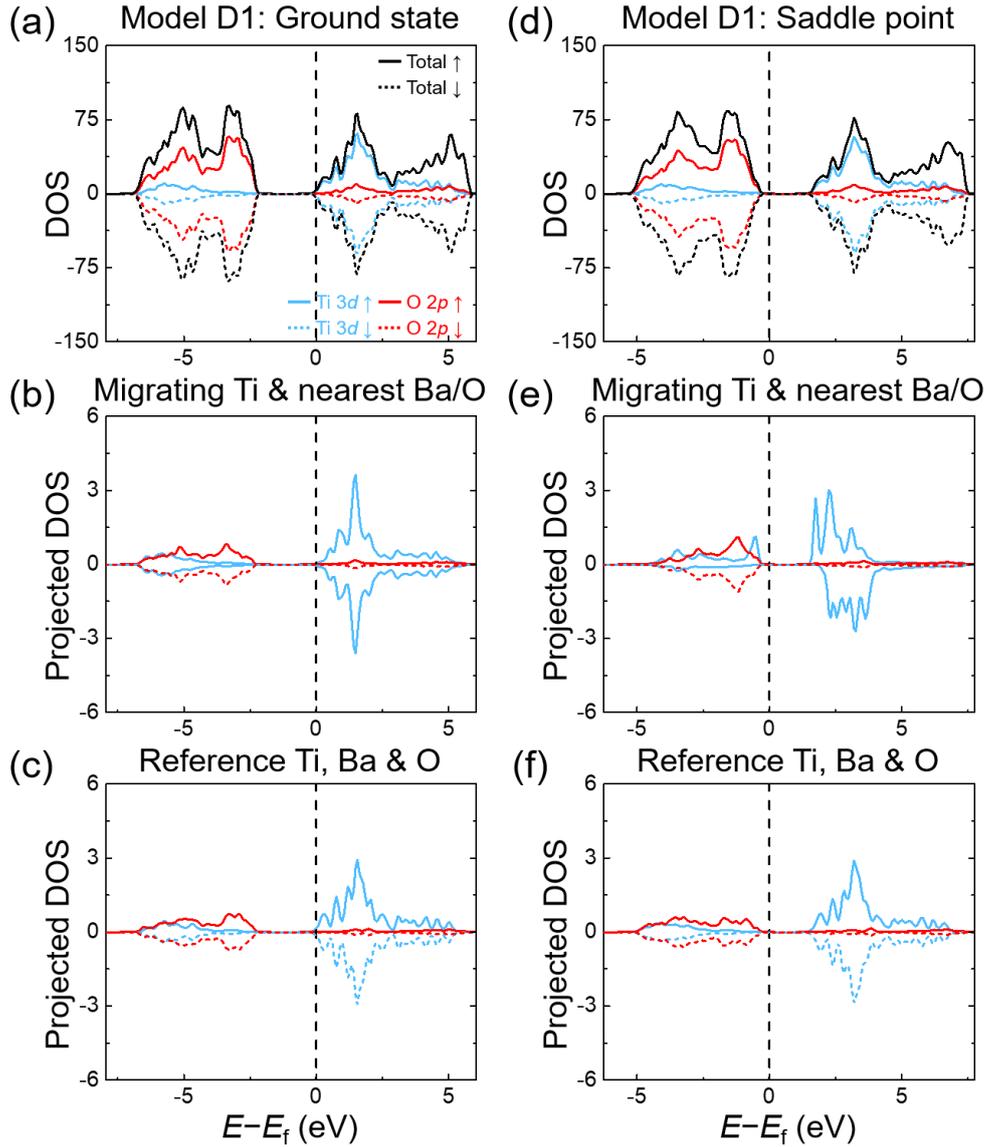

**Figure S16** Calculated density of states (DOS) of cubic BaTiO$_3$ for Ti$^{3+}$ 100 migration **Model D1**. Ground state: (a) total DOS, (b) projected DOS of (to be) migrating Ti (in blue) and nearest O (in red), and (c) projected DOS of non-participating reference Ti (in blue) and O (in red). Saddle-point state: (d) total DOS, (e) projected DOS of migrating Ti (in blue) and nearest O (in red), and (f) projected DOS of non-participating reference Ti (in blue) and O (in red) at the saddle-point configuration. In each figure, Fermi energy is set to be zero and spin-up and spin-down states are plotted as positive and negative, respectively.

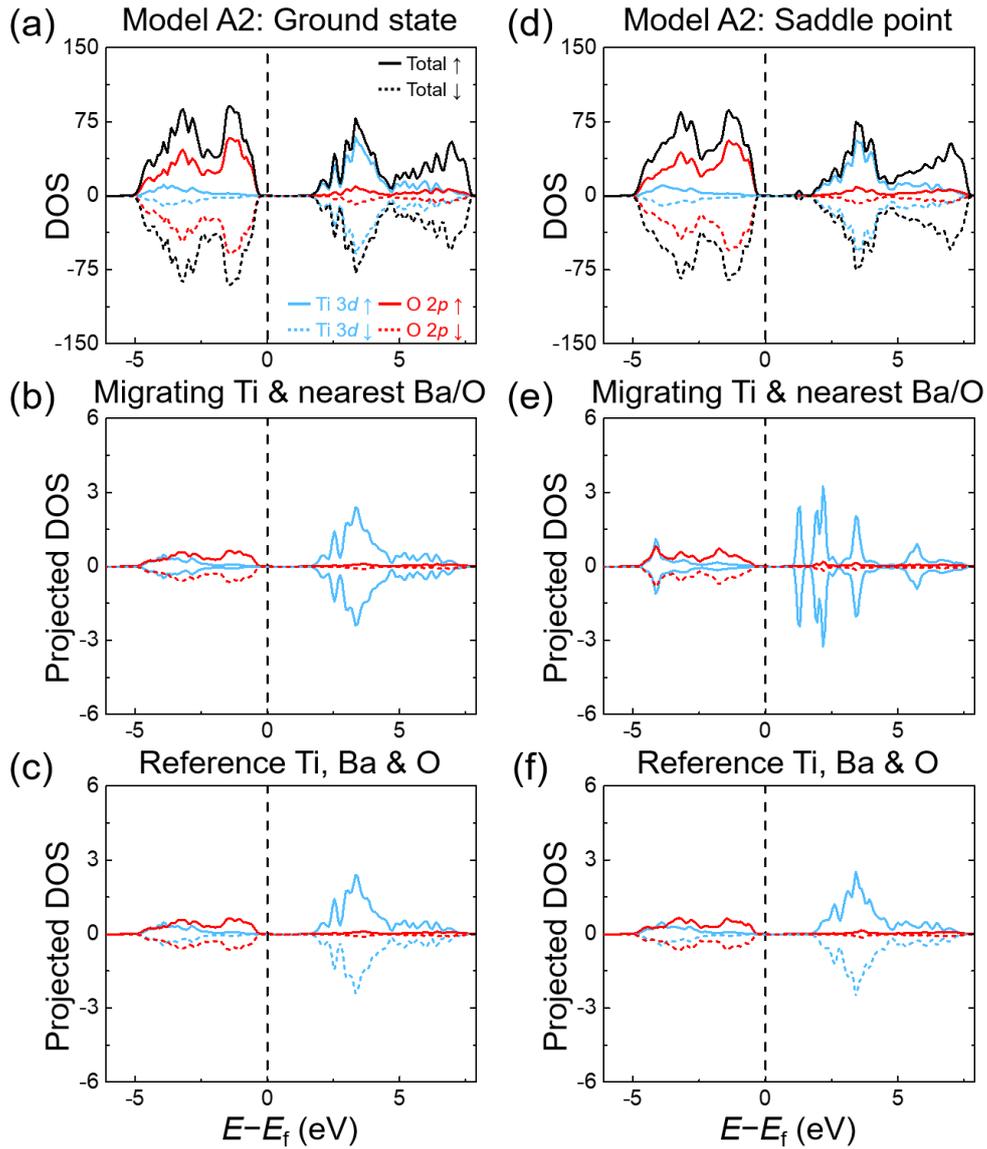

**Figure S17** Calculated density of states (DOS) of cubic BaTiO$_3$ for Ti$^{4+}$ 110 migration **Model A2**. Ground state: (a) total DOS, (b) projected DOS of (to be) migrating Ti (in blue) and nearest O (in red), and (c) projected DOS of non-participating reference Ti (in blue) and O (in red). Saddle-point state: (d) total DOS, (e) projected DOS of migrating Ti (in blue) and nearest O (in red), and (f) projected DOS of non-participating reference Ti (in blue) and O (in red) at the saddle-point configuration. In each figure, Fermi energy is set to be zero and spin-up and spin-down states are plotted as positive and negative, respectively.

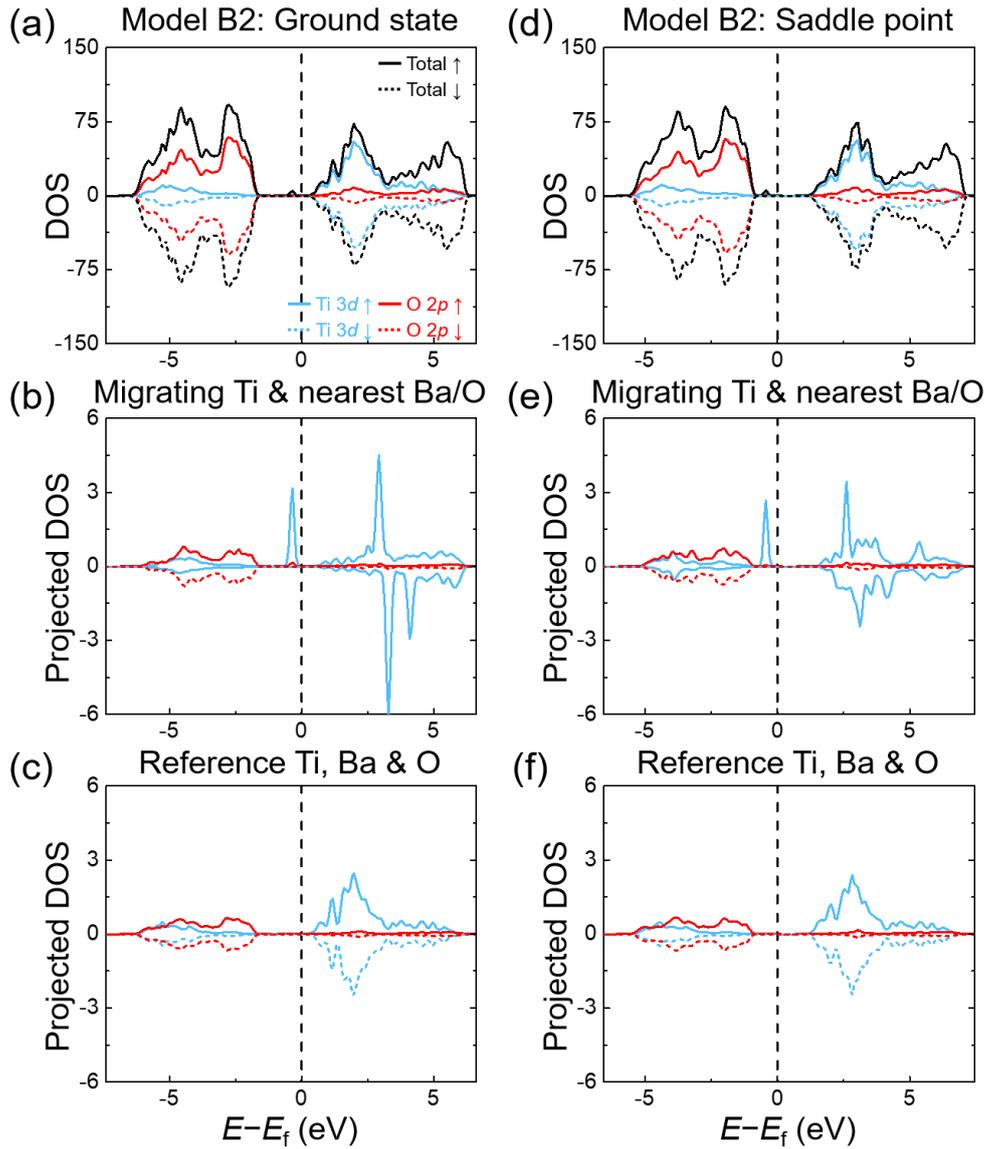

**Figure S18** Calculated density of states (DOS) of cubic BaTiO$_3$ for Ti$^{3+}$ 110 migration **Model B2**. Ground state: (a) total DOS, (b) projected DOS of (to be) migrating Ti (in blue) and nearest O (in red), and (c) projected DOS of non-participating reference Ti (in blue) and O (in red). Saddle-point state: (d) total DOS, (e) projected DOS of migrating Ti (in blue) and nearest O (in red), and (f) projected DOS of non-participating reference Ti (in blue) and O (in red) at the saddle-point configuration. In each figure, Fermi energy is set to be zero and spin-up and spin-down states are plotted as positive and negative, respectively.

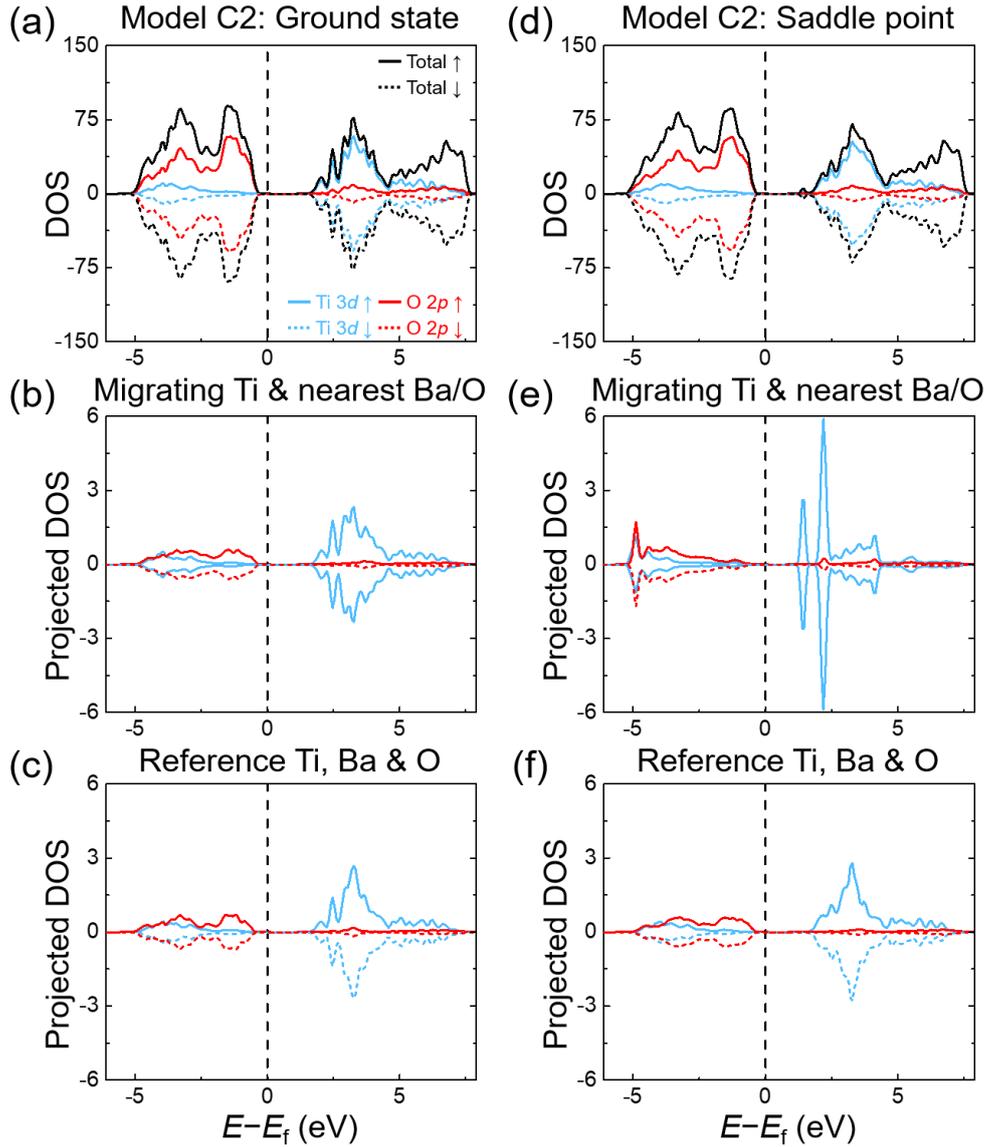

**Figure S19** Calculated density of states (DOS) of cubic BaTiO$_3$ for Ti$^{4+}$ 110 migration **Model C2**. Ground state: (a) total DOS, (b) projected DOS of (to be) migrating Ti (in blue) and nearest O (in red), and (c) projected DOS of non-participating reference Ti (in blue) and O (in red). Saddle-point state: (d) total DOS, (e) projected DOS of migrating Ti (in blue) and nearest O (in red), and (f) projected DOS of non-participating reference Ti (in blue) and O (in red) at the saddle-point configuration. In each figure, Fermi energy is set to be zero and spin-up and spin-down states are plotted as positive and negative, respectively.

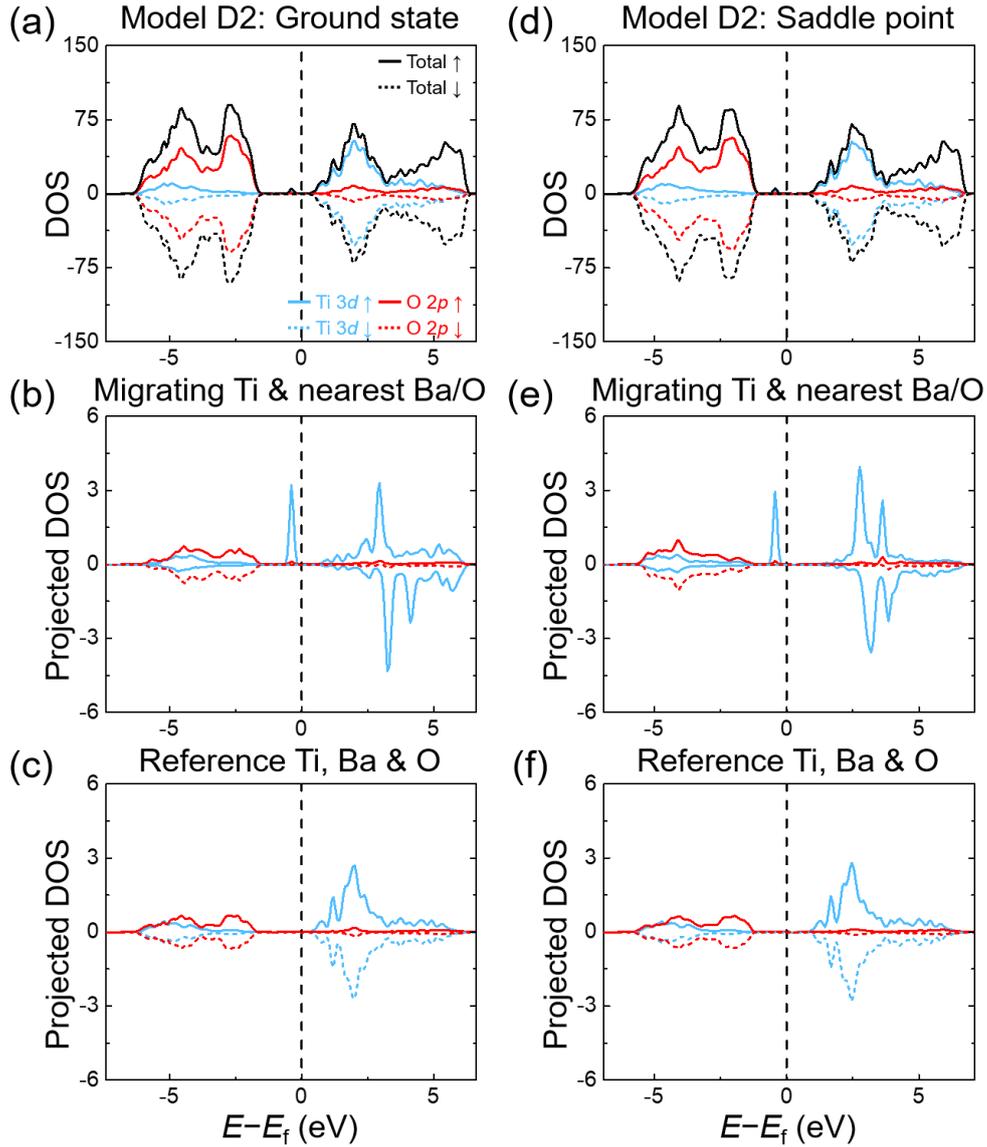

**Figure S20** Calculated density of states (DOS) of cubic BaTiO$_3$ for Ti$^{3+}$ 110 migration **Model D2**. Ground state: (a) total DOS, (b) projected DOS of (to be) migrating Ti (in blue) and nearest O (in red), and (c) projected DOS of non-participating reference Ti (in blue) and O (in red). Saddle-point state: (d) total DOS, (e) projected DOS of migrating Ti (in blue) and nearest O (in red), and (f) projected DOS of non-participating reference Ti (in blue) and O (in red) at the saddle-point configuration. In each figure, Fermi energy is set to be zero and spin-up and spin-down states are plotted as positive and negative, respectively.